\documentclass[%
 reprint,
 amsmath,amssymb,
 aps
]{revtex4-2}

\usepackage{graphicx}
\usepackage{dcolumn}
\usepackage{bm}

\usepackage{orcidlink}
\usepackage{physics}
\usepackage{subcaption}
\usepackage{xcolor}
\usepackage{textcomp}
\usepackage{xstring}

\definecolor{lightgray}{rgb}{0.95,0.95,0.95}

\usepackage{algorithm}
\usepackage{algpseudocode}

\usepackage{cleveref}

\crefformat{equation}{Eq.~(#2#1#3)}
\Crefformat{equation}{Eq.~(#2#1#3)}

\crefformat{figure}{Fig.~(#2#1#3)}
\crefname{algorithm}{Alg.}{Algs.}
\crefformat{algorithm}{Alg.~(#2#1#3)}
\crefname{table}{Table}{Tables}
\crefformat{table}{Table~(#2#1#3)}

\crefname{definition}{Definition}{Definitions}
\crefformat{definition}{Def.~(#2#1#3)}
\crefmultiformat{definition}{Defs.~(#2#1#3)}%
{ and~(#2#1#3)}{, (#2#1#3)}{, and~(#2#1#3)}

\crefrangeformat{equation}{Eqs.~(#3#1#4--#5#2#6)}
\Crefrangeformat{equation}{Eqs.~(#3#1#4--#5#2#6)}

\crefmultiformat{equation}{Eqs.~(#2#1#3)}%
{ and~(#2#1#3)}{, (#2#1#3)}{, and~(#2#1#3)}

\crefname{chapter}{Chapter}{Chapters}
\Crefname{chapter}{Chapter}{Chapters}
\creflabelformat{chapter}{#2#1#3}

\crefformat{appendix}{Appendix~(#2#1#3)}
\crefrangeformat{appendix}{Appendices~(#3#1#4--#5#2#6)}
\crefmultiformat{appendix}{Appendices~(#2#1#3)}{ and~(#2#1#3)}{, (#2#1#3)}{ and~(#2#1#3)}

\usepackage{amsthm}
\theoremstyle{definition}
\newtheorem{definition}{Definition}[section]

\theoremstyle{remark}
\newtheorem*{remark}{Remark}

\newcommand{\GraphicEquationH}[3]{%
    \raisebox{-#1\height}{\includegraphics[height=#2\textheight]{#3}}%
}
\usepackage{mathrsfs}

\newcommand{\tensorA}[2]{A^{#1}_{#2}}
\newcommand{\locT}[1][\ell]{A^{[#1]}}
\newcommand{\locTel}[2]{A^{#1}_{#2}}
\newcommand{\dims}[1]{\left\lceil #1 \right\rceil}

\newcommand{\E}{\mathcal{E}}

\newcommand{\Inner}[2]{ \langle #1, #2 \rangle}
\newcommand{\Stnp}[2]{\mathrm{St} (#1,#2)}
\newcommand{\St}{\Stnp{n}{p}}
\newcommand{\TxM}[2]{T_{#1} \, #2}
\newcommand{\TxStnp}[3]{\TxM{#1}{\Stnp{#2}{#3}}}
\newcommand{\TxSt}[1]{\TxStnp{#1}{n}{p}}

\newcommand{\SVF}{\mathfrak{X}}
\newcommand{\K}{\mathcal{K}}
\newcommand{\X}{\mathcal{X}}

\renewcommand{\L}{\mathcal{L}}
\newcommand{\e}{\mathrm{e}}
\newcommand{\R}{\mathbb{R}}
\newcommand{\I}{\mathbb{I}}
\newcommand{\D}{\mathcal{D}}

\renewcommand{\H}{\mathscr{H}}
\newcommand{\T}{\mathscr{T}}
\newcommand{\fancyL}{\mathscr{L}}
\newcommand{\Choi}{\mathcal{C}}
\newcommand{\N}{\mathbb{N}}
\newcommand{\secref}[1]{\S~\ref{#1}}
\newcommand{\Super}{\mathcal{S}}

\newcommand{\Rank}{\mathrm{R}}
\newcommand{\Xvec}{\mathbf{X}}
\DeclareMathOperator{\gradR}{\mathrm{grad}}
\DeclareMathOperator{\Hess}{\mathrm{Hess}}
\newcommand{\Lket}[1]{\left|{#1}\right\rangle\!\rangle}
\newcommand{\code}[1]{%
    \StrSubstitute{#1}{_}{\textunderscore}[\temp]%
    \texttt{\temp}%
}

\DeclareMathOperator*{\argmin}{arg\,min}
\renewcommand{\O}{\mathcal{O}}

\newcommand{\Xperp}{X_{\perp}}
\newcommand{\Skew}{\mathrm{ skew}}
\newcommand{\Symm}{\mathrm{ sym}}

\newcommand{\Zvec}{\boldsymbol{Z}}
\newcommand{\DOF}{\text{DOF}}

\DeclareMathOperator{\EX}{\mathbb{E}}

\begin{document}

\preprint{APS/123-QED}

\title{A Riemannian Approach to the Lindbladian Dynamics of a Locally Purified Tensor Network}

\author{Emiliano Godinez-Ramirez \orcidlink{0000-0002-8583-3268}}%
\email{cristian.godinez@tum.de}
\affiliation{Technical University of Munich, School of Computation, Information and Technology, Boltzmannstra{\ss}e 3, 85748 Garching, Germany}
\affiliation{Institute for Quantum-Inspired and Quantum Optimization, Hamburg University of Technology, Germany
}
\author{Richard M.~Milbradt \orcidlink{0000-0001-8630-9356}}
\email{r.milbradt@tum.de}
\affiliation{Technical University of Munich, School of Computation, Information and Technology, Boltzmannstra{\ss}e 3, 85748 Garching, Germany}
\author{Christian B.~Mendl \orcidlink{0000-0002-6386-0230}}
\email{christian.mendl@tum.de}
\affiliation{Technical University of Munich, School of Computation, Information and Technology, Boltzmannstra{\ss}e 3, 85748 Garching, Germany}
\affiliation{Technical University of Munich, Institute for Advanced Study, Lichtenbergstra{\ss}e 2a, 85748 Garching, Germany}
\date{\today}

\begin{abstract}
Tensor networks offer a valuable framework for implementing Lindbladian dynamics in many-body open quantum systems with nearest-neighbor couplings. In particular, a tensor network ansatz known as the Locally Purified Density Operator employs the local purification of the density matrix to guarantee the positivity of the state at all times. Within this framework, the dissipative evolution utilizes the Trotter-Suzuki splitting, yielding a second-order approximation error. However, due to the Lindbladian dynamics' nature, employing higher-order schemes results in non-physical quantum channels. In this work, we leverage the gauge freedom inherent in the Kraus representation of quantum channels to improve the splitting error. To this end, we formulate an optimization problem on the Riemannian manifold of isometries and find a solution via the second-order trust-region algorithm. We validate our approach using two nearest-neighbor noise models and achieve an improvement of orders of magnitude compared to other positivity-preserving schemes. In addition, we demonstrate the usefulness of our method as a compression scheme, helping to control the exponential growth of computational resources, which thus far has limited the use of the locally purified ansatz.
\end{abstract}

\maketitle

\section{\label{sec:intro}Introduction}

The study of many-body quantum systems is of utmost importance in various branches of physics, including quantum information and quantum computing. Moreover, to understand realistic physical systems, we ought to consider how these interact with their environment, i.e., we need to understand how they evolve under the influence of noise. This shifts our focus to what is known as open quantum systems \cite{Breuer_Petruccione_2010}. In particular, we are interested in the evolution under Markovian noise, giving rise to the Lindblad master equation \cite{Breuer_Petruccione_2010,Manzano_2020}

Due to the limited analytical methods available to describe such systems, numerical methods play a pivotal role in understanding the so-called \textit{dissipative dynamics}. In this context, tensor networks variational ansätze emerge as an efficient approach capable of describing several systems of interest \cite{Hastings_2006_tensor_networks}. One such structure, known as the Locally Purified Density Operator (LPDO), exploits the main features of tensor networks ans\"atze while preserving the physical properties of open quantum systems \cite{Verstraete_2004_mpo,Cuevas_2013_purification}. Werner et al. \cite{Werner_2016} introduced an algorithm yielding second-order approximation errors based on the ubiquitous Trotter-Suzuki splitting \cite{trotter-higher-splitting} to simulate the Lindblad evolution with nearest-neighbor interactions using the LPDO ansatz. However, due to the semigroup property of Markovian dynamics, it remains an open problem to simulate the dissipative evolution with higher-order accuracy.

In this work, we aim to improve the accuracy of the dissipative simulations under nearest-neighbor Lindbladian dynamics. To this end, we leverage the gauge freedom present in the Kraus representation of quantum channels to formulate an optimization problem on the Riemannian manifold  of isometry matrices, i.e., the Stiefel manifold. We then proceed to solve it via the second-order trust-region algorithm, taking advantage of its global convergence properties \cite{princeton_manifolds}(c.f. \cite{brieger_gst}, which applies a similar approach to Gate Set Tomography). We cover the core concepts needed to understand the main optimization problem addressed in this work in \secref{sec:preliminaries}. Then, we leverage this theory to formulate a real-valued cost function defined on the Stiefel manifold in \secref{sec:formulation}.

Once we formulate the optimization problem and define the Riemannian geometry to address it, we assemble these elements into the main numerical algorithm of this work. In \secref{sec:implementation}, we outline the algorithm, discuss the creation of the optimization ansatz, and introduce the framework for employing our algorithm as a compression scheme.
We conclude this work by benchmarking our approach against other completely positive trace-preserving (CPTP) schemes, evaluating the performance of the compressed representation, and testing the applicability of the optimized operators when used within a larger system. These simulations are presented in \secref{sec:simulations}. For all of them, we have employed two nearest-neighbor models, namely the \textit{Kitaev wire} \cite{kitaev_wire} and a modification of the sparse Pauli-Lindblad model \cite{pec_spl} we call \textit{Pseudo Sparse Pauli-Lindblad} (PSPL). 

\section{Background}\label{sec:preliminaries}

\subsection{Tensor networks and open quantum systems}\label{sec:back-tn-oqs}

In this work, our goal is to simulate the dissipative Lindbladian dynamics of a quantum system governed by the master equation
 \begin{equation}
    \label{eq:lindblad-differential}
    \frac{d \rho}{dt}  = \L(\rho),
\end{equation}
where the density matrix $\rho$ describes the systems's state and the Lindbladian generator $\L$ takes the form
\begin{equation}
        \L(\rho) = \sum_{k} \left( {L}_k {\rho} {L}_k^\dagger - \frac{1}{2} {L}_k^\dagger {L}_k {\rho} - \frac{1}{2} {\rho} {L}_k^\dagger {L}_k \right) \, ,
        \label{eq:lindbladian}
    \end{equation}
with the jump operators $L_k$ modeling the interaction between the system and its environment. The solution of \cref{eq:lindblad-differential} for a finite-dimensional system with initial state $\rho_0 := \rho(t=0)$ is obtained by
\begin{equation}
    \label{eq:lindblad-solution}
    \Lket{\rho} = \Lambda \Lket{\rho_{0}}, \qquad \Lambda = \e^{t \hat{\L}},\qquad t \geq 0,
\end{equation}
where $\Lket{\rho}$ and $\hat{\L}$ are the vectorizations of $\rho$ and $\L$, respectively. The latter vectorized superoperator can be expressed as

\begin{equation}
    \hat{\L} = \sum_k \left[ L_k \otimes L_k^* - \frac{1}{2}(L^\dagger_k L_k \otimes \I + \I \otimes L^T_k L^*_k) \right] \, .
    \label{eq:lindblad-vectorized}
\end{equation}

Due to the limited analytical methods available to describe open quantum systems, numerical methods play a pivotal role in understanding the dissipative dynamics arising from \cref{eq:lindblad-differential}. In this light, tensor networks variational ans\"atze \cite{Hastings_2006_tensor_networks} appear as an efficient approach capable of describing several systems of interest, with the Matrix Product States (MPS) \cite{Fannes_Nachtergaele_Werner_1992_mps,perezgarcia2007matrix} as the archetype class used to study one-dimensional pure states.

On the other hand, one-dimensional mixed states are still a relatively unexplored frontier by tensor network techniques. A first attempt at this employs the class of Matrix Product Density Operators (MPDO) \cite{Verstraete_2004_mpo,Zwolak_2004_mpo}. However, locally verifying the positivity of the global MPDO is NP-hard \cite{Kliesch_positive_np}, so performing local truncations usually destroys the positive definiteness of the state, yielding numerical instabilities in time evolution algorithms. A second approach uses the so-called unraveling of the master equation \cite{Breuer_Petruccione_2010,optics_trajectories} to simulate mixed states by means of MPS methods, together with stochastic quantum trajectories. This approach comes at the cost of a sampling overhead with the number of trajectories. In this work, we will leverage a third structure introduced in \cite{Verstraete_2004_mpo,Cuevas_2013_purification}, known as the Locally Purified Density Operator (LPDO).

\begin{definition}[Locally Purified Density Operator]\label{def:lpdo}
    Given a density matrix $\rho$, its local purification operator $F$ satisfies
    \begin{equation}
        \rho = F F^\dagger,
        \label{eq:lpdo}
    \end{equation}
    and its variational tensor network representation for $N$ sites is
    \begin{equation}
        \begin{aligned}
            \relax[F]^{s_1, \ldots, s_N}_{r_1, \ldots, r_N} =& \!\!\!\!\ \sum_{m_1, \ldots, m_{N-1}} \!\!\!\! \tensorA{[1]s_1,r_1}{m_0, m_1} \tensorA{[2]s_2,r_2}{m_1, m_2} \ldots \tensorA{[N]s_N,r_N}{m_{N-1}, m_{N}} \\
            =& \GraphicEquationH{0.45}{0.09}{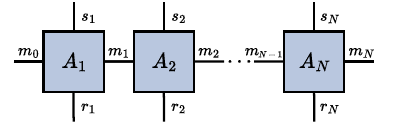} \, .
            \label{eq:lpdo-tensors}
        \end{aligned}
    \end{equation}
    We use the following notation for the tensor legs' dimensions
    \begin{equation}
        \dims{n} := \{1,2,\dots,n\} \, ,
        \label{eq:notation-dimensions}
    \end{equation}
    thus, we can write each rank-4 tensor $\{A^{[\ell]}\}^N_{\ell=1}$ as
    \begin{equation}
        \locT = \left(
            \locTel{[\ell] s_\ell, r_\ell}{m_{\ell-1}, m_\ell}
                \right)_{s_\ell \in \dims{d}, \, r_\ell \in \dims{K}, \, m_{\ell-1} \in \dims{M_{\ell-1}}, \, m_\ell \in \dims{M_\ell}} \, ,
        \label{eq:lpdo-A-tensor}
    \end{equation}
    where $s_\ell$ is the \textbf{physical} index, $r_\ell$ is the \textbf{Kraus} (or \textbf{rank}) index, and $m_\ell$ is the \textbf{bond} index, with dimensions $d$, $K$, and $M$, respectively. For uniformity, the first and last tensors have bond dimensions $m_0 = m_N = 1$. Notice how the representation of $\rho$ in terms of tensors $\{ \locT \}$ is highly non-unique since, in addition to the usual gauge freedom present in MPS along the bond index, this structure has a gauge freedom along the Kraus index, i.e., \cref{eq:lpdo} remains invariant under a unitary transformation of $F$. 
\end{definition}
This work focuses on the scenario where the jump operators $L_k$ act non-trivially only on nearest-neighbor sites $\ell$ and $\ell+1$. Therefore, the Lindbladian superoperator acting on all $N$ sites can be written as
\begin{equation}
    \hat{\L} = \sum_{\ell=1}^{N} \hat{\L}^{[\ell,\ell+1]} \, ,
    \label{eq:full_local_Liouvillian}
\end{equation}
Where each $\hat{\L}^{[\ell,\ell+1]}$ is of the form shown in \cref{eq:lindblad-vectorized}. To simulate the Markovian dynamics of one-dimensional open quantum systems, as given by \cref{{eq:lindblad-solution}}, we employ a practical algorithm described by Werner et al. \cite{Werner_2016} that leverages the positivity-preserving properties of the LPDO structure. Similar to the Time Evolving Block Decimation (TEBD) \cite{Vidal_2003_tebd} scheme, for a small time-step $\tau$, this algorithm splits $\e^{\tau \hat{\L}}$ into several Trotter-Suzuki layers \cite{kliesh_trotter_open} made up of mutually commuting terms. From Lemma 4 in \cite{Werner_2016}, the Trotter-Suzuki approximation for the Markovian dynamics for nearest-neighbors couplings is
\begin{equation}\label{eq:trotter_two_sites}
    \e^{\tau \hat{\L}} = \e^{\tau \hat{\L_o} /2} \e^{\tau \hat{\L_e}} \e^{\tau \hat{\L_o} /2} + O(\tau^3) \, ,
\end{equation}
where the odd and even layers are made up of mutually commuting terms
\begin{equation}\label{eq:Lindblad_even_odd}
    \hat{\L_o} :=  \sum_\ell \hat{\L}^{[2\ell-1,2\ell]} \ \text{ and } \
    \hat{\L_e} := \sum_\ell \hat{\L}^{[2\ell,2\ell+1]}.
\end{equation}
Thus, the exponentials become
\begin{equation}\label{eq:Lindblad_even_odd_exp}
    \e^{\tau \hat{\L_o}} =  \prod_\ell \e^{\tau \hat{\L}^{[2\ell-1,2\ell]}} \ \text{ and } \
    \e^{\tau \hat{\L_e}} = \prod_\ell \e^{\tau \hat{\L}^{[2\ell,2\ell+1]}}.
\end{equation} 
Each superoperator can be seen as a tensor acting on the vectorization of the LPDO in \cref{eq:lpdo}. In other words, they are rank-$4N$ tensors acting on the sites $[1, \ldots, N]$ and their conjugated counterpart $[1*, \ldots, N*]$. Then, using a reshuffling of the tensor legs, which we call the \textit{global-to-local} transformation $G_L$, we obtain the local superoperators
\begin{equation}\label{eq:super-global-local}
    \begin{aligned}
        \e^{\tau \hat{\D}^{[\ell,\ell+1]}} &= \left( \e^{\tau \hat{\L}^{[\ell,\ell+1]}} \right)^{G_L} \\
        &= \GraphicEquationH{0.5}{0.1}{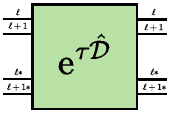} \, .
    \end{aligned}
\end{equation}
See \cref{app:form-global-local} for details on this map and the specific form of the tensors. This transformation allows us to rewrite \cref{eq:Lindblad_even_odd_exp} as
\begin{equation}\label{eq:exp_lindblad_even_odd}
    \e^{\tau \hat{\L}_{o, L}} = \bigotimes_\ell \e^{\tau \hat{\D}^{[2\ell-1,2\ell]}} \ \text{ and } \
    \e^{\tau \hat{\L}_{e, L}} = \bigotimes_\ell \e^{\tau \hat{\D}^{[2\ell,2\ell+1]}} \, ,
\end{equation}
where we added the $L$ subscript to emphasize that these superoperators are a localized version of the original $\hat{\L_o}$ and $\hat{\L_e}$. As explained in Appendix C of \cite{Werner_2016}, to perform the dissipative dynamics within the LPDO framework, we first need to obtain the Kraus representation of the superoperators in \cref{eq:exp_lindblad_even_odd}. To this end, we make use of the following relation between the superoperator and Kraus representations of a quantum channel
\begin{equation}
    \GraphicEquationH{0.49}{0.1}{local_superoperator.pdf} =  \GraphicEquationH{0.5}{0.1}{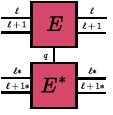},
    \label{eq:super-to-kraus-two-sites}
\end{equation}
with the Kraus tensor
\begin{equation}
    \begin{aligned}
        E^{[\ell,\ell+1]}_q & = \left( E^{[\ell,\ell+1] \, s^{\textrm out}_\ell, s^{\textrm out}_{\ell+1}}_{q, \,  s^{\textrm in}_\ell, s^{\textrm in}_{\ell+1}} \right)_{s^{\textrm out}_\ell, s^{\textrm out}_{\ell+1}, s^{\textrm in}_\ell, s^{\textrm in}_{\ell+1}\in \dims{d}}^{q \in \dims{\Rank}} \\
        & = \GraphicEquationH{0.4}{0.12}{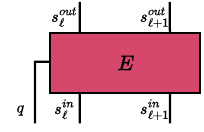} \, .
    \end{aligned}
    \label{eq:kraus-two-site-tensor}
\end{equation}
where $q$ is the Kraus rank index and $\Rank$ is its dimension, i.e., the rank of the channel. The procedure to obtain the Kraus tensor of a quantum channel from its superoperator representation is outlined in \cite{wood2015tensor}.

\begin{figure}[htb]
    \centering
     \subfloat[]{\includegraphics[width=0.35\textwidth]{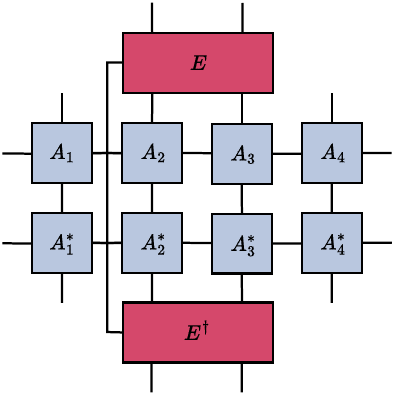}\label{fig:lpdo-kraus-two-sites-01}}
    \qquad
    \subfloat[]{\includegraphics[width=0.35\textwidth]{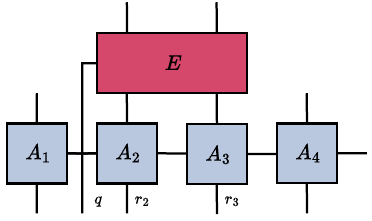}\label{fig:lpdo-kraus-two-sites-02}}
    \qquad
    \caption{Markovian dynamics of a four-site mixed state in LPDO form. Panel (a) shows the action of a quantum channel on the LPDO tensor network. Panel (b) shows the Kraus tensor $E$ being contracted with sites two and three of $F$.}
    \label{fig:lpdo-kraus-two-sites}
\end{figure}

The dissipative evolution using the local Kraus tensors is depicted in \cref{fig:lpdo-kraus-two-sites}. First, in \cref{fig:lpdo-kraus-two-sites-01}, we look at the action of a quantum channel acting on sites $[\ell, \ell+1] = [2,3]$ of a density matrix in its LPDO form for a system with $N=4$. From this representation, we realize the action of the quantum channel on the purification operator $F$ is equivalent to contracting the Kraus tensor $E^{[\ell,\ell+1]}_q$ into the local tensors $A^{[\ell]}$ and $A^{[\ell+1]}$ while merging the channel index $q$ with the variational Kraus indices $r_\ell$ and $r_{\ell+1}$, as shown in \cref{fig:lpdo-kraus-two-sites-02}. By assuming translation invariance and periodic boundary conditions, we arrive at the following diagrammatic relation
\begin{equation}
   (\e^{\tau \hat{\L_o} /2} \e^{\tau \hat{\L_e}} \e^{\tau \hat{\L_o} /2}) \Lket{\rho} \implies \GraphicEquationH{0.5}{0.175}{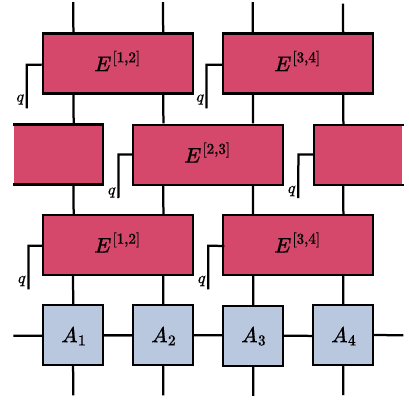} \, .
    \label{eq:approximation-diagram}
\end{equation}

\subsection{Riemannian optimization}\label{sec:back-riemannian}

In \secref{sec:formulation}, we translate our problem into the framework of Riemannian geometry to leverage Riemannian optimization techniques. Thus, we now introduce the key elements of this mathematical formalism necessary to understand the problem formulation.

\begin{definition}[Stiefel manifold]\label{def:stiefel-manifold}
    The \textit{real} Stiefel manifold \cite{princeton_manifolds} is the set
    \begin{equation}
        \St = \{ X \in  \R^{n \times p}: X^T X = \I_p \}, \quad p \leq n,
        \label{eq:stiefel-manifold}
    \end{equation}
    i.e., the set of real isometries or orthonormal matrices.
\end{definition}
To generalize the concept of the gradient of a function defined on the Stiefel manifold, we look at the \textit{tangent space}. 

\begin{definition}[Tangent space]\label{def:tangent-space-first}
    At a certain point $X \in \St$ on the Stiefel manifold, the tangent space $\TxSt{X}$ is a vector space containing the directions we can move in tangent to the manifold. These directions are known as the tangent vectors $Z \in \TxSt{X}$. The tangent space can be parametrized as
    \begin{equation}
        \TxSt{X} = \{ Z \in \R^{n \times p}: Z^T X + X^T Z  = 0 \} \, ,
        \label{eq:tangent-stiefel}
    \end{equation}
    i.e., $X^T Z$ is skew-symmetric \cite{edelman1998geometry_stiefel}.
\end{definition}
The Riemannian gradient for a real-valued function defined on the Stiefel manifold, $f: \St \to \R$, can be evaluated using the projection onto the tangent space $\pi_{T}$ as  
\begin{equation}
    \gradR f(X) = \pi_{T} ( G_X^{-1} \gradR \overline{f}(X)) \, ,
    \label{eq:gradient-metric}
\end{equation}
where $G_X^{-1}$ is the inverse metric operator, and  $\gradR \overline{f}(x)$ is the \textit{euclidean gradient} of $f$ \cite{nguyenoperatorvalued_metric_based}.
Similarly, to generalize the directional derivative, and with it the Hessian, to a Riemannian manifold, we look at the Riemannian connection $\nabla$.
\begin{definition}[Riemannian connection]\label{def:riemannian-connection}
    The Riemannian connection $\nabla$ is a map connecting the different tangent spaces across the manifold, allowing us to differentiate a vector field $\hat{Z}$ along another vector field $\hat{W}$. That is, at a point $X \in \St$, it is the map
    \begin{equation}
        \begin{aligned}
            \nabla&: \TxSt{X} \times \SVF(X) \to \SVF(X) \\
            &: (W, \hat{Z}) \to \nabla_W \hat{Z} \, ,
        \end{aligned}
        \label{eq:riemannian-connection-abstract}
    \end{equation}
    where $\SVF(X)$ is the set of vector fields on $\St$ whose domain includes $X$ \cite{princeton_manifolds}. As shown in Theorem 3.1 of \cite{nguyenoperatorvalued_metric_based}, the form of $\nabla$ is metric-dependent. See \cref{app:imp-metric} for details on this dependence.
\end{definition}
Then, the Riemannian Hessian (vector product) of the function $f$ at $X$ is the operator $\Hess f(X)$ sending a tangent vector $Z \in T_X \St$ to the tangent vector
\begin{equation}
    \Hess f(X)[Z] = \nabla_{Z} \gradR f (X) \, .
    \label{eq:hessian-riemannian}
\end{equation}
where $\gradR f$ is the Riemannian gradient from \cref{eq:gradient-metric} \cite{princeton_manifolds}. Analogous to the Euclidean case, we can use the Riemannian gradient and Hessian to find the next iterate in an optimization scheme. However, following a tangent vector, generally, does not keep us in $\St$. Thus, we ought to use a retraction.
\begin{definition}[Retraction]\label{def:retraction}
    This update step is realized up to first-order \cite{zhang2020newton_retraction} by the smooth retraction map 
    \begin{equation}\label{eq:retraction-def}
        R_X: \TxSt{X} \to \St: Z \to R_X(X + Z) \, .
    \end{equation}
\end{definition}

For conciseness, we define $R_X(Z):= R_X(X + Z)$. Using the retraction, we can map an optimization problem defined on the Stiefel manifold onto the tangent space using the pullback $\hat{f}_X$
\begin{equation}
    \hat{f}_{X} : T_{X} \St \to \R : Z \to f(R_{X}(Z)).
    \label{eq:trust-pullback}
\end{equation}
Then, we can define a model $\hat{m}_X$ as the Taylor expansion of $\hat{f}_X$ around the origin element $0_X$ in the tangent space $T_X \St$. In particular, we look at the quadratic model \cite{princeton_manifolds}
\begin{equation}
    \begin{aligned}
        \hat{m}_X(Z) &= f(X) + \Inner{\gradR f(X)}{Z}_X \\
        &+ \frac{1}{2}\Inner{\Hess f(X)[Z]}{Z}_X , \quad Z \in \TxSt{X} \, .
    \end{aligned}
    \label{eq:model-first-order}
\end{equation}
To minimize $\hat{m}_X$, and consequently $f$,  we use the Trust region algorithm as it combines the local superlinear rate of convergence of the Newton method with global convergence properties \cite{princeton_manifolds}. Starting at $X_\beta$, we obtain the next optimization iterate $X_{\beta+1}$ by finding an approximate solution $Z_\beta$ to the minimization subproblem
\begin{equation}
    \begin{aligned}
        \min_{Z \in T_{X_\beta} \St} &\hat{m}_{X_\beta}(Z) = f(X_\beta) + \Inner{\gradR f(X_\beta)}{Z}_{X_\beta}\\
        &+ \frac{1}{2}\Inner{\Hess f(X_\beta)[Z]}{Z}_{X_\beta} , \,  \Inner{Z}{Z}_{X_\beta} \leq \Delta^2 \, ,
    \end{aligned}
    \label{eq:trust-subproblem}
\end{equation}
where $\Delta$ is known as the \textit{trust-region radius}. The next iterate is then $X_{\beta+1} = R_{X_\beta}(Z_\beta)$. In this work, we choose to solve the trust-region subproblem approximately by means of a computationally light algorithm known as the conjugate-gradient (CG) method (see 7.3.2 of \cite{princeton_manifolds} for details). To assess the quality of the model, we define the quotient
\begin{equation}
    \rho_\beta = \frac{f(X_\beta) - f(R_{X_\beta}(Z_\beta))}{\hat{m}_{X_\beta}(0_{X_\beta}) - \hat{m}_{X_\beta}(Z_{\beta})} = \frac{\hat{f}_{X_\beta}(0_{X_\beta}) - \hat{f}_{X_\beta}(Z_{\beta})}{\hat{m}_{X_\beta}(0_{X_\beta}) - \hat{m}_{X_\beta}(Z_{\beta})} \, .
    \label{eq:trust-quotient}
\end{equation}
Based on the value of $\rho_\beta$ we identify the following cases:
\begin{itemize}
    \item If $\rho_\beta \ll 1$, the model is very inaccurate: we \textbf{reject} $R_{X_\beta}(Z_\beta)$ and \textbf{reduce} $\Delta$.
    \item If $\rho_\beta$ is small but not as drastically: we \textbf{accept} $R_{X_\beta}(Z_\beta)$ but \textbf{reduce} $\Delta$.
    \item If $\rho_\beta \approx 1$, the model and function have a good agreement: we \textbf{accept} $R_{X_\beta}(Z_\beta)$ and \textbf{increase} $\Delta$.
    \item If $\rho_\beta \gg 1$, the model is inaccurate, but (luckily) the cost function is decreasing: in this case, we can try our luck and \textbf{increase} $\Delta$ to hope for further decrease in $f$.
\end{itemize}

\section{Formulation}\label{sec:formulation}

In \secref{sec:back-tn-oqs}, we discussed how to describe a mixed state as a Locally Purified Density Operator (LPDO) tensor network ansatz. In particular, we saw how to simulate the dissipative dynamics of nearest-neighbor Lindblad operators, which yields the second-order Trotter approximation
\begin{equation*}
    \e^{\tau \hat{\L}} = \e^{\tau \hat{\L_o} /2} \e^{\tau \hat{\L_e}} \e^{\tau \hat{\L_o} /2} + O(\tau^3) \, .
\end{equation*}

However, finding higher-order Trotter approximations for Markovian dynamics remains an open problem \cite{Werner_2016}. This challenge arises from the semigroup property of Lindblad superoperators since, contrary to the unitary evolution, the negative coefficients in higher-order Trotter splittings \cite{trotter-higher-splitting,YOSHIDA1990262,BLANES2002313} would result in non-physical maps. In this work, we aim to address this limitation by leveraging the gauge freedom of quantum channels along with the Riemannian optimization methods on the Stiefel manifold we described in \secref{sec:back-riemannian}. Our objective is to achieve a better approximation error for the dissipative dynamics of nearest-neighbor Lindbladians while maintaining the alternating structure of the layers shown by the second-order Trotter splitting in \cref{eq:approximation-diagram}.
\subsection{From superoperators to isometries}\label{sec:form-superop-isometries}
As we saw in \cref{eq:super-global-local}, we can transform our expressions to focus on the local superoperators $\e^{\tau \hat{\D}^{[\ell,\ell+1]}}$. Then, we can employ these superoperators to reformulate our problem to fit into a Riemannian optimization scheme. Using the Kraus operators' completeness relation
\begin{equation}
    \label{eq:kraus-completeness}
    \sum_{q=1}^{R} E_q^\dagger E_q = \I \, ,
\end{equation}
we can follow the procedure outlined in \cite{Luchnikov_2021} to cast these operators into the Stiefel manifold $\St$ we introduced in \cref{def:stiefel-manifold}. 
We will label the whole superoperator-isometry transformation as $\mathrm{S_T}$, and define the map
\begin{equation}
    \mathrm{S_T} : \T(\H_1, \H_2) \to \St : \Lambda \to X = (\Lambda)^{\mathrm{S_T}} \, ,
    \label{eq:super-to-stiefel}
\end{equation}
where $\T(\H_1, \H_2)$ is the space of linear maps from operators in the Hilbert space $\H_1$, to operators in the Hilbert space $\H_2$, see \cref{def:superoperators} of the appendix. The inverse map from isometries to superoperators
\begin{equation}
    \mathrm{S_P} = (\mathrm{S_T})^{-1}
    \label{eq:stiefel-to-super}
\end{equation}
follows immediately from this definition.

\subsection{Defining the cost function: approximation error}\label{sec:form-cost-function}

Now that our operators belong to the Stiefel manifold, we can construct a real-valued cost function $f$ on this manifold, i.e.,
\begin{equation}
    f: \St \to \R \, .
\end{equation}
To this end, we look at the difference between the approximation given by the right-hand side of \cref{eq:trotter_two_sites} and the full superoperator on the left-hand side. In particular, we use the computationally relevant Frobenius norm induced by the Euclidean inner product
\begin{equation}
    \norm{   \e^{\tau \hat{\L}} -   \e^{\tau \hat{\L}}_{\text{approx}}} \, .
    \label{eq:cost-function-general}
\end{equation}
It is customary to divide the time evolution up to a final time $\tau$ into smaller time steps of size $\Delta t$, such that
\begin{equation}
    \tau = n_\tau \cdot \Delta t \, ,
    \label{eq:timesteps}
\end{equation}
with $n_\tau \in \N$ the number of time steps. If we use the second-order Trotter splitting from \cref{eq:trotter_two_sites} for each $\Delta t$, we arrive at the following pattern for the approximate superoperators
\begin{align}
    \e^{\tau \hat{\L}}_{\text{approx}} & = \GraphicEquationH{0.475}{0.09}{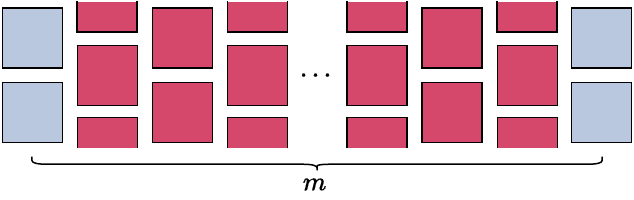} \label{eq:even-odd-multisteps-simplified}
\end{align}
where the \textit{blue} blocks correspond to the terms with half-time steps, while the \textit{red} blocks correspond to the full time steps. The number of layers $m$ in \cref{eq:even-odd-multisteps-simplified} is
\begin{equation}
    m = 3n_\tau - (n_\tau-1) = 2n_\tau + 1 \, .
    \label{eq:number-of-layers}
\end{equation}
We see the final form of the splitting with $n_\tau$ time steps showcases the same odd-even structure from \cref{eq:trotter_two_sites}. However, due to this new partition structure, the approximation error in terms of $\Delta t$ becomes
\begin{equation}
    \e^{\tau \hat{\L}} = \e^{\tau \hat{\L}}_{\text{approx}} + \mathcal{O}(\Delta t^2) \, .
    \label{eq:splitting-error-n-time-steps}
\end{equation}

Finally, in order to arrive at the final expression of the cost function, we lay out the assumptions used:
\begin{itemize}
    \item To achieve translational symmetry, we assume the operators $\hat{\D}^{[\ell, \ell+1]}$, and therefore the isometries $X$, are the same for all $\ell \in [1, N]$. 
    \item We use the splitting for $n_\tau$ time steps to simulate the evolution up to time $\tau$ with $m$ \textit{distinct} layers. Contrary to the repeating structure from the original Trotter splitting, we choose all the layers in our ansatz to be independent of each other to increase the degrees of freedom of the optimization in hopes of achieving a higher expressivity.
    \item We restrict our optimization to the \textit{real} Stiefel manifold to avoid dealing with non-holomorphic functions defined on complex manifolds. 
\end{itemize}

The composition of superoperators is achieved through usual matrix multiplication. Then, by labeling each of the isometries on the $\alpha$-th layer as $X_{\alpha}$, and $\{X\}_\alpha$ as the list of $m$ isometries, we can express the cost function as
\begin{equation}
    f(\{X\}_\alpha) = \norm{ \e^{\tau \hat{\L}} - \prod_\alpha^m \left(\bigotimes_\ell (X_{\alpha})^{\mathrm{S_P}} \right)^{L_G}} \, ,
    \label{eq:cost-function-stiefel}
\end{equation}
where $\mathrm{S_P}$ is the isometry-superoperator map from \cref{eq:stiefel-to-super} and $L_G$ is the local-to-global transformation, i.e., the inverse of the global-to-local map we introduced in \cref{eq:super-global-local} (see \cref{eq:local-to-global-map} for details). We can define the vector of isometries
\begin{equation*}
    \Xvec := (X_1, \ldots, X_m) \, ,
\end{equation*}
and with it, the superoperator 
\begin{equation*}
    \Super(\Xvec) = \prod_\alpha^m \left(\bigotimes_\ell (X_{\alpha})^{\mathrm{S_P}} \right)^{L_G} \in \R^{d^{2N} \times d^{2N}}
\end{equation*}
acting on the full system. Thus, the main optimization problem of interest in this work can be formulated as
\begin{equation}
    \min_{\Xvec} f(\Xvec) = \norm{ \e^{\tau \hat{\L}} - \Super(\Xvec)} \, .
    \label{eq:cost-function-xvec}
\end{equation}
In other words, we want to find the list of isometries that minimize the approximation error of the dissipative dynamics.

\section{Algorithm}\label{sec:implementation}

So far, we have covered the underlying theory required for this work and used it to formulate a real-valued cost function defined on the Stiefel manifold. It is now time to tackle the optimization problem. The main algorithm \textit{A Riemannian Approach to Lindbladian Dynamics} used to solve \cref{eq:cost-function-xvec} is outlined in \cref{alg:riemannian-approach-main}. The detailed construction of the Riemannian gradient, Hessian, and retraction, as well as the choice of the Riemannian metric, are out of the scope of this work but can be found in \cref{app:imp-metric,app:imp-retraction,app:imp-gradient,app:imp-hessian}. Instead, in this section, we discuss the creation of the ansatz employed in our optimization scheme and introduce the degree of freedom of the Kraus tensors that allows us to use our algorithm as a compression scheme.

\begin{figure}
\begin{minipage}{\linewidth}
\begin{algorithm}[H]
	\caption{A Riemannian Approach to Lindbladian Dynamics}\label{alg:riemannian-approach-main}
    \textbf{Input:} \\
    $L_k$: List of Lindblad operators acting on nearest-neighbors\\
    $\tau$: Final time of Lindblad dissipative evolution \\
    $n_\tau$: Number of time steps for Trotter splitting \\
    $N$: Number of sites in the LPDO structure \\
    $d$: Physical dimension per site in the LPDO \\
    $\Rank$: Desired rank of Kraus tensor acting on nearest-neighbors in the LPDO tensors \\
    $I$: Number of iterations for trust-region optimization \\
    \textbf{Output:} \\
    $f_\beta$: Cost function history, evaluated after every iteration $\beta$ of the optimization\\
    $\tilde{\Xvec}$: List of isometries minimizing approximation error after $I$ iterations
	\begin{algorithmic}[1]
        \Require All $L_k \in \R^{d^2 \times d^2}$
        \State $\e^{\tau \hat{\L}} \gets$ Create reference superoperator from $\{L_k\}_k$ and \cref{eq:full_local_Liouvillian}. \Comment{Assume translational symmetry for all $N$ sites.}
        \State $\{\Lambda_\alpha\}_\alpha \gets$ Create $m = 2n_\tau + 1$ superoperators using the Trotter splitting from \cref{eq:trotter_two_sites} and the decomposition in \cref{eq:even-odd-multisteps-simplified}.
        \State $ \Xvec^0 \gets$ Generate ansatz of isometries from superoperators $\{\Lambda_\alpha\}_\alpha$.
        \State $f( \Xvec^{\beta}, \e^{\tau \hat{\L}}) \gets$ Define cost function taking as input the exact superoperator $\e^{\tau \hat{\L}}$ and list of isometries $\Xvec^{\beta}$.
        \State $\gradR f \gets$ Define Riemannian gradient from cost function.
        \State $R_X \gets$ Define retraction taking a tangent space vector $Z$ back to Stiefel manifold.
        \State $\Hess f \gets$ Define Riemannian Hessian from cost function and Riemannian gradient.
        \For{$\beta \gets 0$ \textbf{to} $I$ $-1$}
            \State $\Xvec^{\beta+1} \gets$ Perform an iteration of the trust-region \\
            \hspace{5em} algorithm with $(f,\gradR f, \Hess f, R_X,\Xvec^{\beta})$.
        \EndFor
    \end{algorithmic}
\end{algorithm}
\end{minipage}
\end{figure}

The entirety of this work has been implemented in \code{Python} \cite{python311} and can be found open-sourced on GitHub as a package called \code{OpenTN} \cite{opentn}. As a backend for manipulating the tensors we have used \code{NumPy} \cite{harris2020array} and \code{JAX} \cite{jax2018github}. The latter has also been utilized due to its automatic differentiation capabilities.

\subsection{Creating the ansatz: Compressed factorization}\label{sec:imp-ansatz}

As for many other gradient-based optimization algorithms, the initial optimization point, i.e., the ansatz, can play an important role in the number of iterations needed in the optimization scheme. Sometimes, it can even determine whether the algorithm reaches a local minimum or not. In this work, we choose to start at the even-odd splitting given by the second-order Trotter from \cref{eq:trotter_two_sites}. We have seen that for $n_\tau$ time steps, this ansatz results in $m$ layers made up of local superoperators
\begin{equation}
    \Lambda_\alpha \in \R^{d^4 \times d^4}, \quad \forall \alpha \in \dims{m}\, .
    \label{eq:superop-local-ansatz}
\end{equation}

We can then convert each superoperator into its isometry representation using the $\mathrm{S_T}$ map from \cref{eq:super-to-stiefel}.The Cholesky decomposition of the Choi matrix is a crucial step of the $\mathrm{S_T}$ transformation, see \cref{app:cholesky-decomp} for details. The degree of freedom in the choice of the channel rank $\Rank$ used in this factorization is of particular importance
\begin{equation}
    \Choi_{\Rank} = \sum_{q=1}^{\Rank} E^{\, q}_{s^{out}, s^{in}} (E^{\, q}_{s^{out}, s^{in}})^\dagger \, ,
    \label{eq:imp-choi-r-decomposition}
\end{equation}
with $s^{out}, s^{in} \in \dims{d^2}$, and $\Choi_{\Rank}$ the rank-$\Rank$ approximation of the Choi matrix $\Choi$.  The obvious choice is to set $\Rank = \rank(\Choi)$ so that $\Choi_\Rank = \Choi$. We call this the \textit{natural Choi rank} of the channel. However, we can arbitrarily increase this number up to $\Rank_{max} = d^4$ or compress it down to $\Rank_{min}=1$. The larger the Choi rank $\Rank$, the higher the expressivity of each layer in the ansatz, generally resulting in a lower approximation error. However, this increase in expressivity comes at the cost of an increase in the complexity of the optimization and, consequently, the computational cost. Another crucial factor to take into account is the effect of the channel rank on the growth of the Kraus dimension in the LPDO structure, where the indices $r_\ell$ get contracted with the channel index $q$ of dimension $\Rank$, as shown in \cref{fig:lpdo-kraus-two-sites}. In \secref{sec:sim-rank-up}, we explore how the choice of $\Rank$ affects our optimization algorithm and discuss the benefits of using a compressed representation of the quantum channels. 

Once we choose a rank $\Rank$, we apply the $\mathrm{S_T}$ transformation on the list of superoperators $\{\Lambda_\alpha\}_{\alpha\in\dims{m}}$ to obtain a list of isometries
\begin{equation}
     \Xvec^0 := (X^0_1, \ldots, X^0_m), \quad X_{\alpha} \in \R^{n \times p}, \quad \forall \alpha \in \dims{m}\, ,
    \label{eq:isometries-ansatz}
\end{equation}
where we have defined $n = \Rank d^2$ and $p = d^2$. The superscript $0$ indicates that this corresponds to the zero-th iteration of the optimization scheme. The numerical implementation of the Cholesky decomposition is discussed in  \cref{app:cholesky-decomp}.

\subsection{Parametrizing the tangent space}\label{sec:imp-parametrizing-tangent-space}
To further understand the effect of the channel rank in the optimization scheme, it is useful to look at the number of parameters present in the ansatz. In other words, we want to understand how $\Rank$ affects the expressivity. To this end, we look at the parametrized definition of the tangent space of the Stiefel manifold. The in-detail derivation following the outline from \cite{unitary_constraints} is included in \cref{app:imp-parametrization-sm}.
\begin{definition}[Tangent space - revisited]\label{def:imp-tangent-space}
    Given an element $X \in \St$ of the Stiefel manifold, the tangent space $T_X \St$ at $X$ is
    \begin{equation}\label{eq:imp-tangent-space-param}
        \begin{aligned}
            T_X \St &= \big\{ Z \in \R^{n \times p}: Z = XA + \Xperp B, A \in \R^{p \times p}, \\
            & A + A^T = 0, B \in \R^{(n-p)\times p} \big\} \, .
        \end{aligned}
    \end{equation}
    If we compare \cref{eq:imp-tangent-space-param,eq:tangent-stiefel}, we realize the product $X^T Z$ in the former is exactly the skew-symmetric matrix $A$ of the latter, for $Z \in T_X \St$.
\end{definition}
The arbitrary matrix $B$ in \cref{eq:imp-tangent-space-param} is made of 
$p(n-p)$ independent parameters, while the skew-symmetric matrix $A$ is completely defined by its $p(p-1)/2$
upper diagonal elements. Thus, as a byproduct of the parametrized formulation, we can verify the degrees of freedom of $T_X \St$ to be
\begin{equation}
    \frac{p(p-1)}{2} + p(n-p) = np - \frac{p(p+1)}{2}\, .
    \label{eq:imp-tangent-dimensions}
\end{equation}
Then, the total degrees of freedom present in our $m$-layer ansatz are
\begin{equation}\label{eq:imp-dof-ansatz}
    \begin{aligned}
     \DOF &= \frac{m p}{2}\left( 2n - p - 1 \right) \\
     & = \frac{m d^2}{2}\left( 2 \Rank d^2 - d^2 - 1 \right)\, .
    \end{aligned}
\end{equation}

\section{Simulations}\label{sec:simulations}

In this chapter, we study the performance of the \textit{Riemannian Approach to Lindbladian Dynamics} as introduced in \cref{alg:riemannian-approach-main}. As a testbed, we focus on two models of nearest-neighbor dissipative interactions, with real-valued jump operators, as our optimization is defined on the real Stiefel manifold. Namely, at all neighboring sites $[\ell,\ell+1]$ of an $N$-sites quantum system we consider the Lindblad operators given by the models

\begin{enumerate}
    \item Kitaev wire:
    \begin{equation}
        L_1 = \frac{\sqrt{\gamma}}{4}(a^{\dagger} \otimes \I + \I \otimes a) \, ,
        \label{eq:sim-kitaev}
    \end{equation}
    where $a^{\dagger}$ and $a$ are the two-level quantum harmonic oscillator creation and annihilation operators, respectively, and $\I$ is the $2 \times 2$ identity matrix, as introduced in \cite{kitaev_wire}.
    \item Pseudo Sparse Pauli-Lindblad (PSPL):
    \begin{equation}
        L_k = \sqrt{\gamma}(P_k - Q_k), \quad k \in [\kappa] \, ,
    \end{equation}
    where $P_k, Q_k$ are elements of the Pauli-2 group, such that the total number of Lindblad operators $\kappa \ll 4^2 - 1$, i.e., the Lindblad terms are sparse. Compared to the original sparse Pauli-Lindblad model introduced in \cite{pec_spl}, here we have the difference of two Pauli operators instead of a single Pauli. In particular, in this work, we choose to use two Lindblad operators with weight-one Pauli terms
    \begin{equation}
        \begin{aligned}
            L_1 &= \sqrt{\gamma}(X \otimes \I - \I \otimes X) \, , \\
            L_2 &= \sqrt{\gamma}(Z \otimes \I - \I \otimes Z) \, .
        \end{aligned}
        \label{eq:sim-pspl}
    \end{equation}
\end{enumerate}

For both models, the prefactor $\gamma$ corresponds to the noise strength, which we normalize to $\gamma = 1$ over all simulations. Instead, we vary the final evolution time $\tau$ over the experiments, since this is equivalent to increasing the noise strength, so we can verify the validity of our results across different noise regimes. We have implicitly assumed the local Hilbert space dimension to be $d = 2$, i.e., we work with a system of qubits. We perform the simulations using the first non-trivial even number of sites $N = 4$ yielding an effective Hilbert space dimension $d^{2N} = 2^8$, as this keeps the computation tractable. In the following sections, we test our algorithm under the two aforementioned noise models, while exploring the influence of the channel rank $\Rank$, introduced in \secref{sec:implementation}. Further, we compare the performance of our algorithm against the second-order Trotter splitting \cite{Werner_2016} and a structure-preserving scheme of arbitrary order introduced in \cite{cao2021structurepreserving}. To end this chapter, we leverage the translational invariance and PBC assumed in our ansatz to test the optimized isometries when embedding them into larger systems.

\subsection{Rank up}\label{sec:sim-rank-up}
A fundamental element in our optimization scheme was introduced in \secref{sec:imp-ansatz}: the rank R of the ansatz. This quantity arises in the Cholesky decomposition from \cref{eq:imp-choi-r-decomposition} and determines the column dimension of the Stiefel manifold $p = \Rank d^2$. In this section, we will examine the conjectures presented in \secref{sec:imp-ansatz}, regarding the impact of the ansatz's rank on the performance of the Riemannian algorithm. To this end, we carry out the optimization scheme under the Kitaev and PSPL noise models, while choosing the ansatz's rank $\Rank$ as
\begin{equation}
  \Rank \in \{2, \ldots, \Rank_{\Choi}, \ldots, d^4\}  \, ,
\end{equation}
where $\Rank_{\Choi} = \rank(\Choi)$ is the natural Choi rank, as introduced in \cref{eq:imp-choi-r-decomposition}. We do not use $\Rank=1$, as this would correspond to a unitary evolution.

\begin{figure*}[htb]
    \centering
    \includegraphics[width=0.65\textwidth]{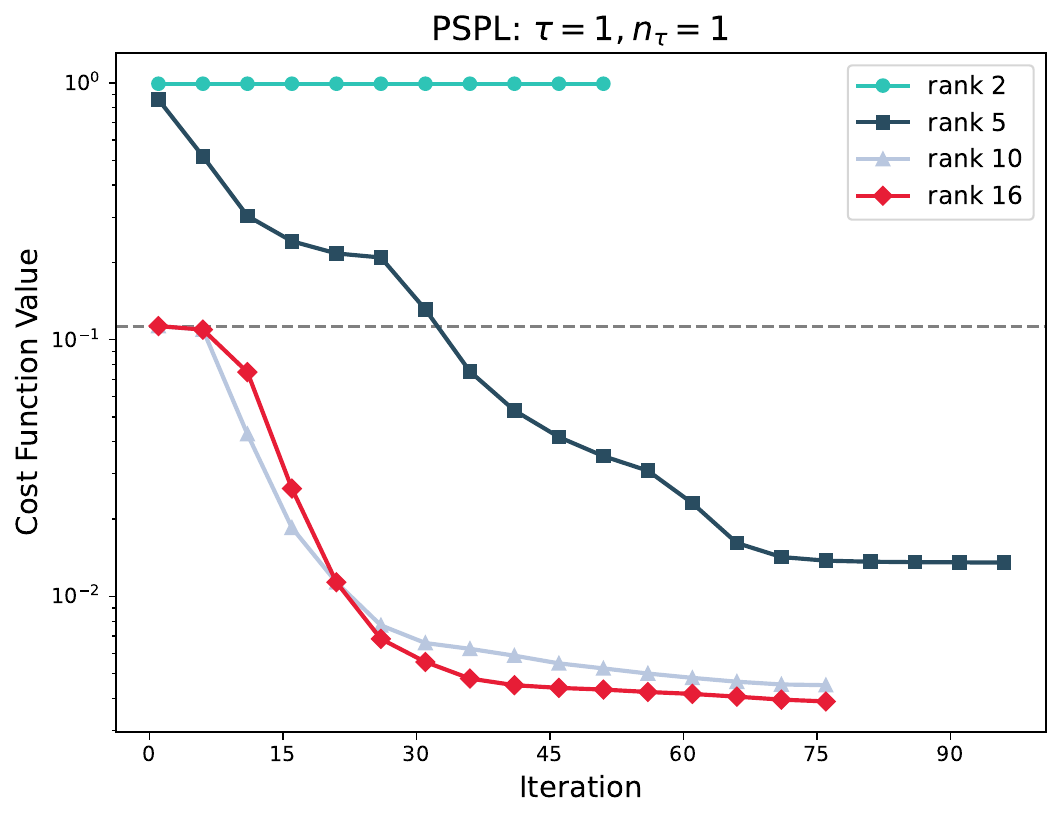}
    \caption{Riemannian optimization with PSPL Lindblad model for a final time of $\tau=1$ and $n_\tau=1$ time steps. The four curves correspond to the use of different ranks for the ansatz. The gray dashed line marks the value of the Trotter splitting error.}
    \label{fig:sim-rank-pspl}
\end{figure*}

 We transform the local superoperator arising from the PSPL Lindblad operators in \cref{eq:sim-pspl} into its Choi representation, to obtain the natural Choi rank of the channel
 \begin{equation*}
    \Rank^{\mathrm{PSPL}}_{\Choi} = 10 \, .
 \end{equation*}
 In other words, the PSPL quantum channel can be described exactly with $10$ Kraus operators. We then perform the Riemannian optimization for a final time of $\tau=1$ with $n_\tau=1$ time steps and visualize the optimization trajectory in \cref{fig:sim-rank-pspl} using four different ranks
 \begin{equation*}
    \Rank \in \{2, 5, 10, 16\} \, .
 \end{equation*}

 As expected, this comparison illustrates how the algorithm's performance improves with increasing rank. The highest ranks $\Rank \in \{10, 16\}$ start the optimization at the Trotter splitting value, marked by the gray dashed line, and exhibit a continuous decrease with the iterations. Due to the increased parameter space, the optimization with $\Rank = 16$ attains the lowest cost function value out of all. However, a more surprising result is yielded by the curve with $\Rank = 5 < \Rank^{\mathrm{PSPL}}_{\Choi}$. Since this ansatz corresponds to a \textit{compressed} representation of the channel, it initially performs worse than the higher rank approximations, as we expected. Yet, after 30 iterations it already reaches the Trotter splitting value, and after 100 iterations, the cost function sees an improvement of nearly one order of magnitude with respect to the Trotter splitting and two orders of magnitude with respect to its ansatz value. This demonstrates that using this compressed representation not only enables a decrease in the approximation error of the dissipative dynamics but also helps control the exponential increase in the Kraus dimensions of the LPDO tensors after each dissipative step. Thus, we can directly employ our algorithm to alleviate the computational cost associated with the noisy evolution of the LPDO structure, cf. \cite{Cheng_2021,müller2024enabling}.

 The other end of the spectrum is shown by the turquoise curve with $\Rank=2$, where the improvement yielded by the optimization is negligible. To understand this behavior, we can look at the Choi rank of the superoperator acting on all $N$ sites arising from the rank-R ansatz, $\Super^{\Rank}(\Xvec)$. The Choi rank of $\Super^{\Rank}(\Xvec)$ is upper bounded as
\begin{equation}
    \Rank_N := \rank \left\{ \left(\Super^{\Rank}(\Xvec)\right)^{R_r} \right\} \leq \Rank^{2m} \, .
    \label{eq:sim-upper-bound}
\end{equation}
To make this expression more tangible, we calculate numerically the Choi rank of the exact full superoperator $\e^{\tau \hat{\L}}$ and compare it to the rank of $\Super^{\Rank}(\Xvec)$ for each of the ansatz ranks in our simulation.

 \begin{table}
    \begin{subtable}[h]{0.2\textwidth}
        \renewcommand{\arraystretch}{1.5}
        \begin{tabular}{c | c| c}
        \hline\hline
        $\Rank$ & $\Rank_N$ & $\Rank_N^{\mathrm{opt}}$\\
        \hline
        Exact & $2^8$ & - \\
        \hline
        2 & 53 & 53\\
        5 & $2^8$ & $2^8$ \\
        10 & $2^8$ & $2^8$ \\
        16 & $2^8$ & $2^8$\\
        \hline
        \end{tabular}
        \caption{Comparison for the PSPL model.}
        \label{tab:ranks-pspl}
    \end{subtable}
    \begin{subtable}[h]{0.2\textwidth}
        \renewcommand{\arraystretch}{1.5}
        \begin{tabular}{c | c | c}
        \hline\hline
        $\Rank$ & $\Rank_N$ & $\Rank_N^{\mathrm{opt}}$ \\
        \hline
        Exact & 45 & - \\
        \hline
        2 & 45 & 117 \\
        4 & 45 & 205\\
        8 & 45 & 94 \\
        16 & 45 & 202\\
        \hline
        \end{tabular}
        \caption{Comparison for the Kitaev model.}
        \label{tab:ranks-kitaev}
    \end{subtable}
    \caption{Comparison of full Choi ranks for different approximations using the two toy models before and after optimization.}
    \label{tab:ranks}
\end{table}

\cref{tab:ranks-pspl} shows both the initial rank $\Rank_N$ and the rank after the optimization $\Rank_N^{\mathrm{opt}}$. From this, we observe that, unlike the higher-rank approximations, the ansatz with $\Rank=2$ reaches only $\approx 1/4$ of the full rank of $\e^{\tau \hat{\L}}$, even after the attempt to optimize it. This result agrees with \cref{eq:sim-upper-bound} and provides an explanation for the limited performance shown in \cref{fig:sim-rank-pspl}.

\begin{figure}[htb]
    \centering
    \includegraphics[width=0.5\textwidth]{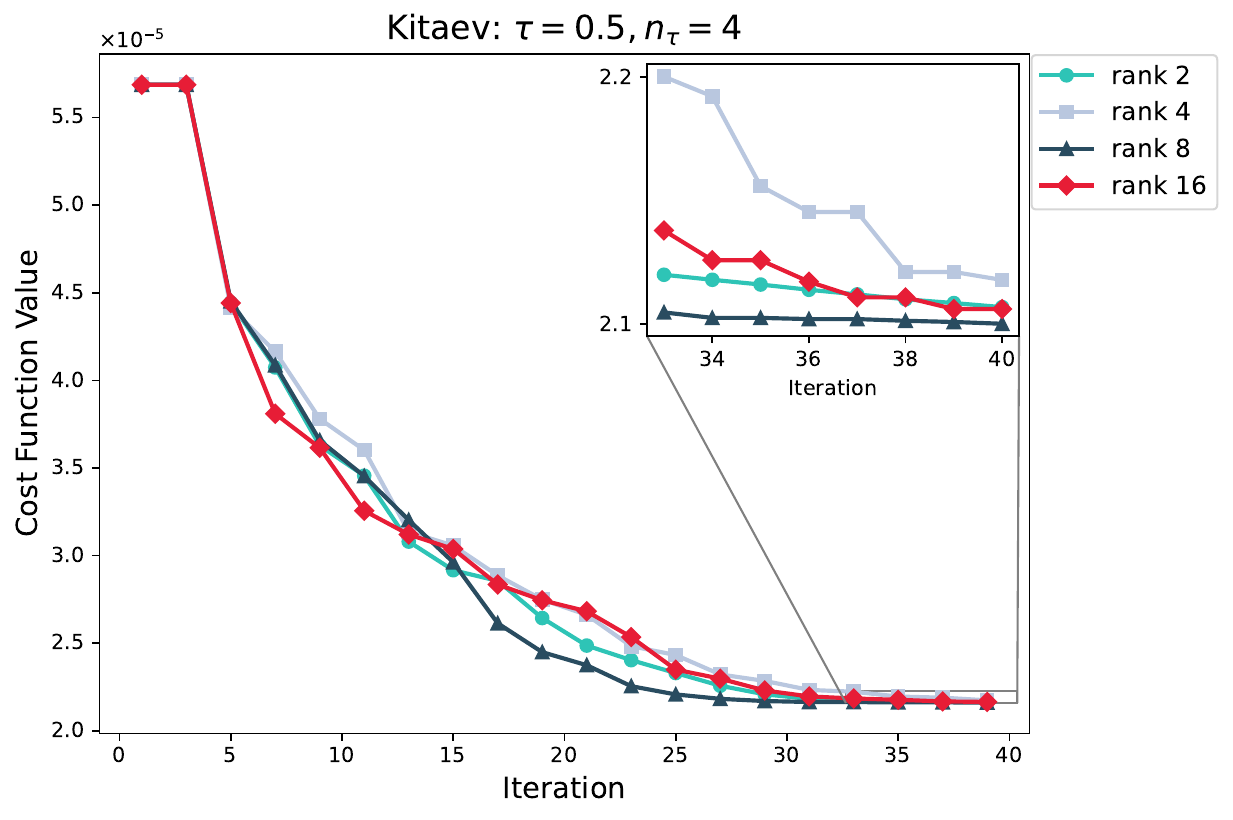}
    \caption{Riemannian optimization with Kitaev Lindblad model for a final time of $\tau=0.5$ and $n_\tau=4$ time steps. The four curves correspond to the use of different ranks for the ansatz. The inset shows a zoom-in version of the last $10$ iterations to emphasize the difference between the ans\"atze.}
    \label{fig:sim-rank-kitaev}
 \end{figure}

Contrary to what we initially expected based on \cref{fig:sim-rank-pspl}, increasing the rank of the ans\"atze does not always lead to considerably better results. An example of this is the simulation under the Kitaev model shown in \cref{fig:sim-rank-kitaev}.
For this computation, we use $\tau = 0.5$, $n_\tau = 4$, and choose $\Rank$ from  $\{2, 4, 8, 16\}$. The first thing to notice is the natural Choi rank, which for this model  is
\begin{equation*}
    \Rank^{\mathrm{Kitaev}}_\Choi = 2 \,.
\end{equation*}
For this reason, all the ans\"atze in this simulation start at the Trotter splitting value. While we notice a consistent decrease in the cost function for all the curves in \cref{fig:sim-rank-kitaev}, it is also noteworthy that all of them reach a plateau at only 40 iterations, and none of them attain a cost function lower than 30\% of the original value. Analogously to the computation for the PSPL model, we calculate the Choi rank of $\Super^{\Rank}(\Xvec)$ before and after the optimization, and show the results in \cref{tab:ranks-kitaev}.

This comparison highlights a couple of things. First, the exact superoperator can be represented by a relatively low rank, $\Rank_N = 45$, which is already achieved by all $\Super^{\Rank}(\Xvec)$ even before the optimization. This can serve as an explanation for the low approximation error at which we start the optimization scheme, hence making it harder to improve upon. Second, in an attempt to improve on this approximation, the rank of all the ans\"atze $\Rank_N^{\mathrm{opt}}$ increases considerably. This would result in a rapid increase of the Kraus dimensions of the LPDO structure without necessarily achieving a substantial decrease in the approximation error of the dissipative dynamics.

\subsection{Riemannian to the test}\label{sec:sim-alg-comparison}

\begin{figure*}[htb]
    \centering
    \includegraphics[width=0.65\textwidth]{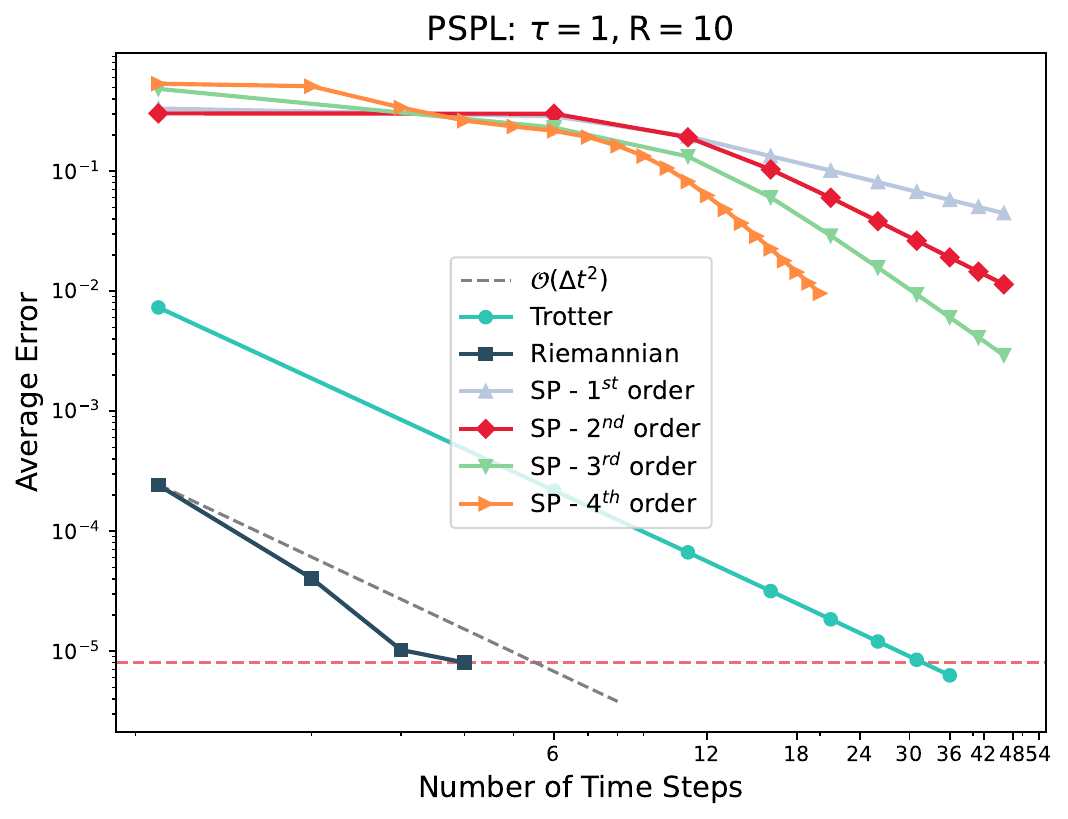}
    \caption{Average error as a function of the number of time steps for different CPTP methods under the PSPL noise model, final time of $\tau=1$, and ansatz rank of $\Rank = 10$. We use the normalized version of the structure-preserving schemes of order $1,2,3$, and 4. The red horizontal line is used to emphasize the number of time steps needed by the Trotter splitting to achieve the same accuracy as the Riemannian optimized method.}
    \label{fig:sim-benchmark-pauli}
 \end{figure*}

To benchmark the performance of our algorithm, we compare its scaling against the second-order Trotter splitting from \cref{eq:trotter_two_sites} and the structure-preserving (SP) scheme introduced by Y. Cao and J. Lu in \cite{cao2021structurepreserving}. In the latter, the authors introduce a family of unnormalized completely positive schemes of arbitrary order. To achieve a fair comparison against the SP methods, we use the normalized version of the schemes and compute the composite superoperator up to fourth order due to the computational cost of higher-order schemes. To this end, we compute the average error as
\begin{equation}
    \EX\left(  \norm*{ \e^{\tau \hat{\L}}(\rho_0) - \Super_{n_\tau} (\rho_0) }  \right) \, ,
    \label{eq:sim-average-error}
\end{equation}
where $\Super_{n_\tau}$ is the superoperator from $n_\tau$ time steps of either the normalized SP schemes, the Trotter splitting, or our Riemannian algorithm. The expectation value is taken over $500$ randomly generated density matrices $\rho_0$. We perform our simulations using the PSPL noise model for a final time of $\tau=1$, ansatz rank of $\Rank = 10$, and plot the average error against the number of time steps $n_\tau$ in \cref{fig:sim-benchmark-pauli}.

The results in \cref{fig:sim-benchmark-pauli} highlight the significant improvement attained by our optimization scheme, decreasing the error for every $n_\tau$ by more than one order of magnitude, compared to the Trotter splitting. As emphasized by the red horizontal line, the Trotter splitting would need $n_\tau > 30$ time steps to achieve the same accuracy obtained by only $n_\tau = 4$ using our method. In addition, comparing our method against the second-order curve showcases how, at some intervals, the Riemannian optimized error follows an order higher than two. This difference becomes more significant as we increase the number of time steps. We attribute this behavior to the increased expressivity of the ansatz since, as shown in \cref{eq:imp-dof-ansatz}, the curve with $n_\tau=4$ corresponds to $m = 9$ layers of isometries and $1350$ degrees of freedom we can optimize over, compared to the $450$ parameters available in the optimization using $n_\tau=1$.

These results also showcase that while the structure-preserving schemes are capable of reaching approximations of arbitrarily high order, the error prefactor is larger than those for the Trotter and Riemannian methods. Moreover, the error order of these methods is only observed for a higher number of time steps. Both of these behaviors are expected and explained in detail in \cite{cao2021structurepreserving}.

\begin{figure}[htb]
    \centering
    \includegraphics[width=0.5\textwidth]{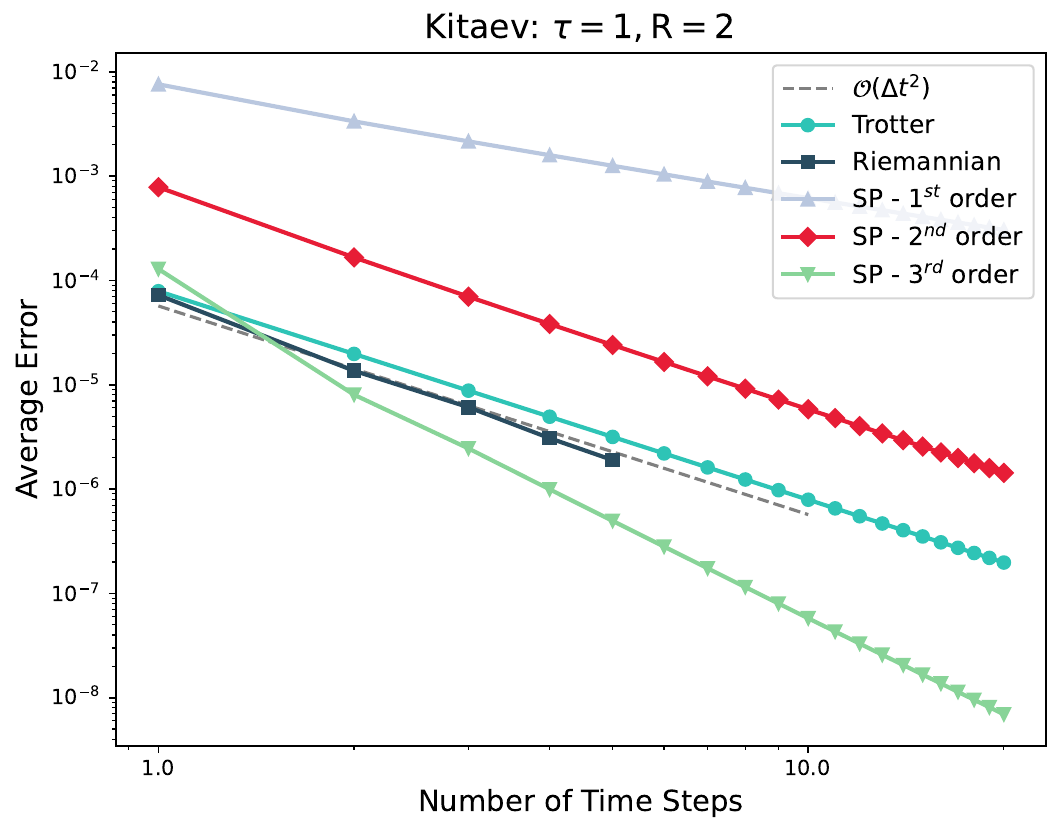}
    \caption{Average error as a function of the number of time steps for different CPTP methods under the Kitaev model, with $\tau = 1$, and $\Rank = 2$, plotted against increasing number of time steps. We use the normalized version of the structure-preserving schemes of order $1,2$, and $3$.}
    \label{fig:sim-kitaev-comparison}
\end{figure}

To end this section, we repeat the above comparison using the Kitaev model, final evolution time $\tau = 1$, and ansatz rank $\Rank = 2$. As shown in \cref{fig:sim-kitaev-comparison}, in this setting, the improvement of the Riemannian optimization compared to the Trotter error is not as significant as for the PSPL model, even though at some intervals, the optimized curve again follows an order slightly higher than two. As previously discussed, we also observe that the three SP schemes employed suffer from a high error prefactor. However, the order-three SP scheme quickly catches up due to the higher approximation order. For values of $n_\tau$ close to $10$, it attains an improvement of orders of magnitude compared to the Trotter splitting. It is, however, worth highlighting that even in scenarios where the structure-preserving schemes outperform the Trotter and Riemannian approximations, they cannot be implemented directly on the LPDO ansatz, as they do not preserve the local structure of the tensors.

For additional insight into the comparison between the superoperators achieved by each scheme, in \cref{app:ranks-schemes}, we compare the Choi ranks $R_N$ for each of the approximation methods using both noise models.

\subsection{Increasing the system size}\label{sec:sim-larger-sites}

In \secref{sec:form-cost-function}, we discussed the assumptions we made when defining our cost function, one of which was the translational invariance of our ansatz. Additionally, we assumed all odd layers to follow periodic boundary conditions, as shown in \cref{eq:even-odd-multisteps-simplified}. Due to these assumptions, although we performed the optimizations for systems of size $N=4$, it is possible to employ the optimized isometries within larger systems. In this section, we use the optimized layers from $N=4$ to verify how well they approximate the dissipative dynamics of a system of $N=6$ sites with an effective dimension of $d^{2N}=2^{12}$.

\begin{figure}[htb]
    \centering
    \subfloat[Kitaev]{\includegraphics[width=0.475\textwidth]{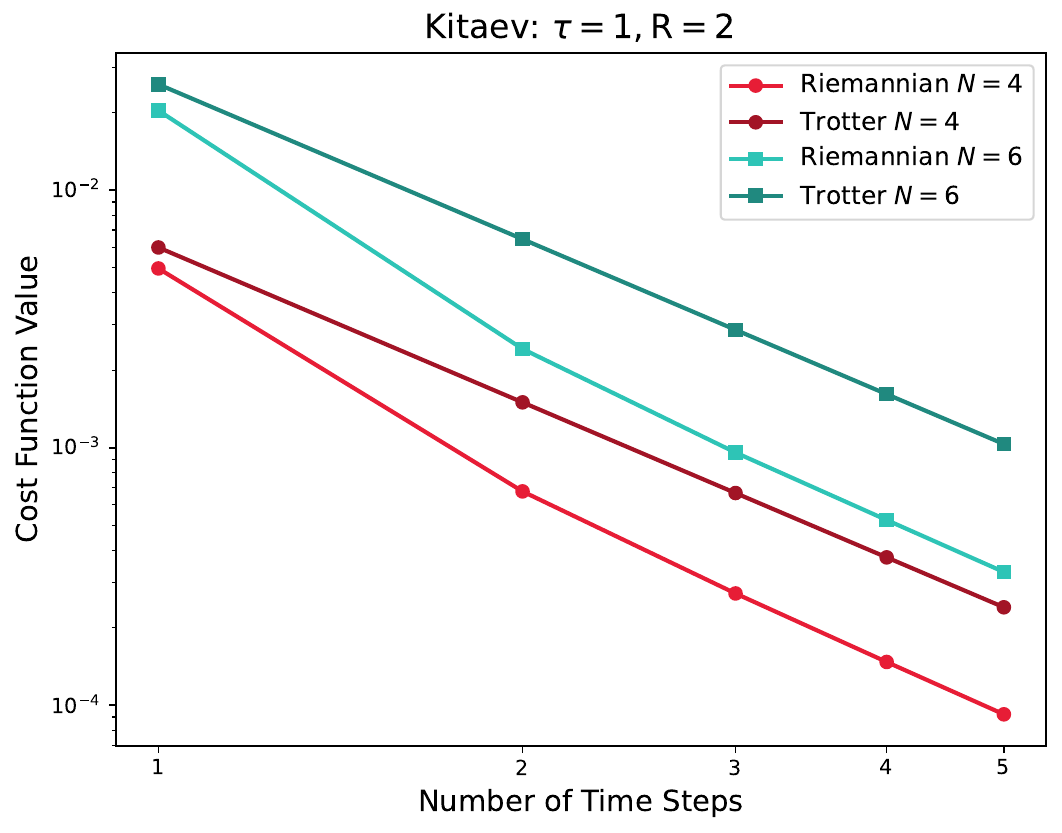}\label{fig:sim-higher-sites-kitaev}}
    \qquad
    \subfloat[PSPL]{\includegraphics[width=0.475\textwidth]{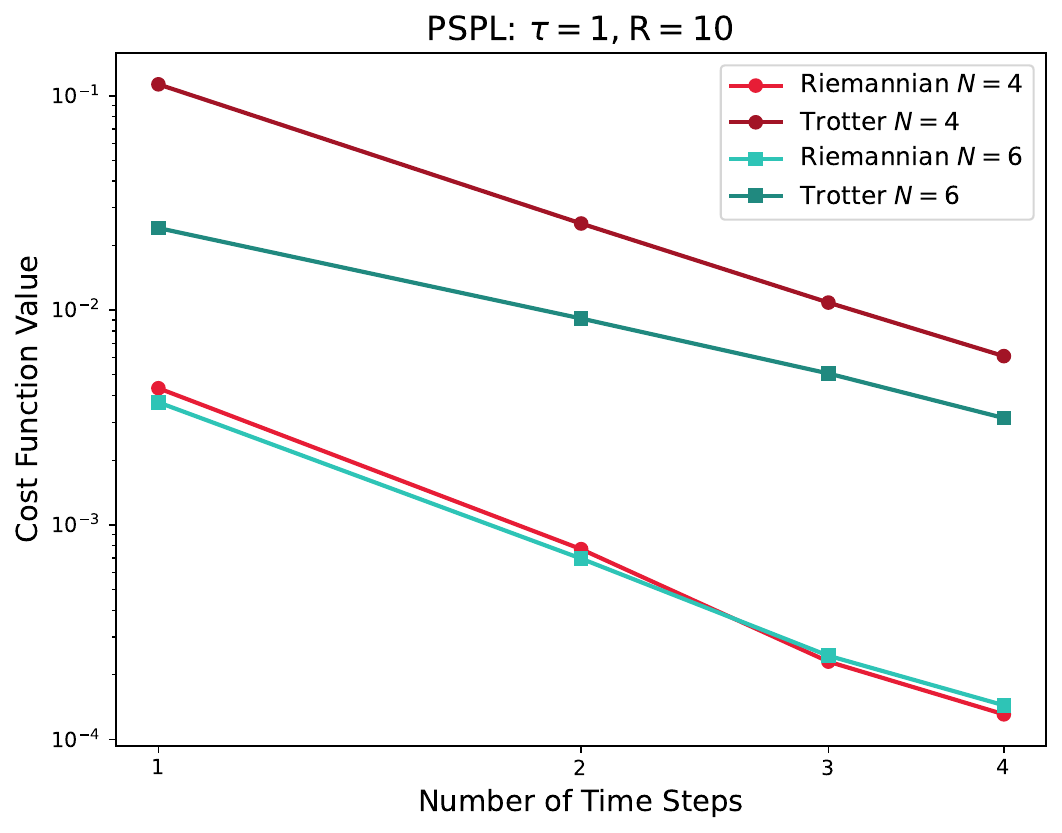}\label{fig:sim-higher-sites-pspl}}
    \caption{Comparison of the cost function from Trotter and Riemannian schemes against the number of time steps $n_\tau$, with evolution time $\tau = 1$, for system sizes $N=4$ and $N=6$. The layers for the Riemannian scheme with $N=6$ use the same isometries obtained with $N=4$. Panel (a) shows the comparison for the Kitaev model with $\Rank=2$, while panel (b) shows the comparison using the PSPL model with $\Rank = 10$.}
    \label{fig:sim-higher-sites}
\end{figure}

For this comparison, we use the optimized isometries $\tilde{\Xvec}$ obtained in \secref{sec:sim-alg-comparison} under the Kitaev and PSPL models, using $\tau = 1$ and their natural ranks, $\Rank=2$ and $\Rank=10$, respectively. \cref{fig:sim-higher-sites} shows the approximation error against the number of time steps $n_\tau$ for the original $N=4$ and the increased size $N=6$. First, in \cref{fig:sim-higher-sites-kitaev} we examine the Kitaev model comparison, from which we observe that the error achieved by the optimized isometries within the larger system closely follows the original approximation curve. In other words, although the Trotter error increases with system size, so does the Riemannian error, resulting in very similar curves for both values of $N$. 

On the other hand, \cref{fig:sim-higher-sites-pspl} shows the comparison using the PSPL model, yielding somewhat different results. Here, we observe how the larger systems' Trotter and Riemannian errors are lower. This behavior is more pronounced for the Trotter scheme, as shown by the difference among the darker curves. However, for both methods with $N=6$, we observe that as the number of time steps increases, the slope of the error decreases. From this observation, we can infer that the error for the system with $N=4$ will be lower for higher $n_\tau$. 
Despite this difference, the trajectory for both system sizes follows a similar trend, indicating an improvement of orders of magnitude regardless of the system size.

\section{Conclusion and Outlook}\label{chapter:conlcusion}

In this work, we employed a framework based on the Locally Purified Density Operator tensor network ansatz to formulate the approximation error of the nearest-neighbor Lindbladian dynamics as an optimization problem on the Stiefel manifold. We showed that our method is capable of achieving an improvement of orders of magnitude with respect to other known schemes. In addition, this formulation allowed us to devise a compressed representation of the quantum channels. This result alone is of particular interest as it would allow us to alleviate the exponential increase in the Kraus index of the LPDO when performing noisy evolution while still improving the approximation error.
Finally, we also showed that our method can improve the approximation error of larger system sizes by optimizing computationally tractable systems. This is especially relevant as the dimension of the Lindblad superoperators grows as $d^{2N}$ with the system size $N$.

A potential area for improvement is the wall time needed by our algorithm, with the most computationally expensive step being the construction of the Riemannian Hessian. To address this, we can implement a more efficient Hessian-vector product computation that does not need to construct the full Hessian. We leave this task for future work. Conversely, one might argue that the wall time of our algorithm should not be considered a strict constraint. This is especially true for a time-independent noise model, where our scheme can be seen as a single pre-processing step, entailing a \textit{non-scaling} overhead.

It remains an open task to study the effect of the specific ansatz in the optimization algorithm, as discussed in \cite{kotil2023riemannian}, since this might influence the complexity of the optimization landscape and, therefore, affect the convergence point and speed. A similar open issue is to develop a rigorous mathematical understanding of the dependence of the algorithm's performance on the properties of the noise model. In this light, studying other physically meaningful noise models with high rank would provide further insight.

Our numerical implementation of the Riemannian gradient and Hessian is based on an automatic differentiation software framework, providing us great flexibility when defining the cost function. Therefore, generalizing the ansatz to a non-translational invariant ansatz would be relatively straightforward. Finally, note that we considered only the real Stiefel manifold. This restricts the optimization to real Kraus channels. Thus, a further generalization of this work could encompass the extension to the complex Stiefel manifold. This would require recounting the tangent space dimensions and the correct formulation of the non-holomorphic complex derivatives.

\begin{acknowledgments}
We would like to thank Fiona Fröhler for the discussions regarding the numerical implementation of the structure-preserving schemes. 
The research is part of the Munich Quantum Valley, which is supported by the Bavarian state government with funds from the Hightech Agenda Bayern Plus.
\end{acknowledgments}

\bibliography{apssamp}

\providecommand{\noopsort}[1]{}\providecommand{\singleletter}[1]{#1}%
\begin{thebibliography}{37}%
\makeatletter
\providecommand \@ifxundefined [1]{%
 \@ifx{#1\undefined}
}%
\providecommand \@ifnum [1]{%
 \ifnum #1\expandafter \@firstoftwo
 \else \expandafter \@secondoftwo
 \fi
}%
\providecommand \@ifx [1]{%
 \ifx #1\expandafter \@firstoftwo
 \else \expandafter \@secondoftwo
 \fi
}%
\providecommand \natexlab [1]{#1}%
\providecommand \enquote  [1]{``#1''}%
\providecommand \bibnamefont  [1]{#1}%
\providecommand \bibfnamefont [1]{#1}%
\providecommand \citenamefont [1]{#1}%
\providecommand \href@noop [0]{\@secondoftwo}%
\providecommand \href [0]{\begingroup \@sanitize@url \@href}%
\providecommand \@href[1]{\@@startlink{#1}\@@href}%
\providecommand \@@href[1]{\endgroup#1\@@endlink}%
\providecommand \@sanitize@url [0]{\catcode `\\12\catcode `\$12\catcode
  `\&12\catcode `\#12\catcode `\^12\catcode `\_12\catcode `\%12\relax}%
\providecommand \@@startlink[1]{}%
\providecommand \@@endlink[0]{}%
\providecommand \url  [0]{\begingroup\@sanitize@url \@url }%
\providecommand \@url [1]{\endgroup\@href {#1}{\urlprefix }}%
\providecommand \urlprefix  [0]{URL }%
\providecommand \Eprint [0]{\href }%
\providecommand \doibase [0]{https://doi.org/}%
\providecommand \selectlanguage [0]{\@gobble}%
\providecommand \bibinfo  [0]{\@secondoftwo}%
\providecommand \bibfield  [0]{\@secondoftwo}%
\providecommand \translation [1]{[#1]}%
\providecommand \BibitemOpen [0]{}%
\providecommand \bibitemStop [0]{}%
\providecommand \bibitemNoStop [0]{.\EOS\space}%
\providecommand \EOS [0]{\spacefactor3000\relax}%
\providecommand \BibitemShut  [1]{\csname bibitem#1\endcsname}%
\let\auto@bib@innerbib\@empty
\bibitem [{\citenamefont {Breuer}\ and\ \citenamefont
  {Petruccione}(2010)}]{Breuer_Petruccione_2010}%
  \BibitemOpen
  \bibfield  {author} {\bibinfo {author} {\bibfnamefont {H.-P.}\ \bibnamefont
  {Breuer}}\ and\ \bibinfo {author} {\bibfnamefont {F.}~\bibnamefont
  {Petruccione}},\ }\href@noop {} {\emph {\bibinfo {title} {The theory of open
  quantum systems}}}\ (\bibinfo  {publisher} {Oxford Univ. Press},\ \bibinfo
  {address} {Oxford},\ \bibinfo {year} {2010})\BibitemShut {NoStop}%
\bibitem [{\citenamefont {Manzano}(2020)}]{Manzano_2020}%
  \BibitemOpen
  \bibfield  {author} {\bibinfo {author} {\bibfnamefont {D.}~\bibnamefont
  {Manzano}},\ }\bibfield  {title} {\bibinfo {title} {A short introduction to
  the {L}indblad master equation},\ }\bibfield  {journal} {\bibinfo  {journal}
  {AIP Advances}\ }\textbf {\bibinfo {volume} {10}},\ \href
  {https://doi.org/10.1063/1.5115323} {10.1063/1.5115323} (\bibinfo {year}
  {2020})\BibitemShut {NoStop}%
\bibitem [{\citenamefont {Hastings}(2006)}]{Hastings_2006_tensor_networks}%
  \BibitemOpen
  \bibfield  {author} {\bibinfo {author} {\bibfnamefont {M.~B.}\ \bibnamefont
  {Hastings}},\ }\bibfield  {title} {\bibinfo {title} {Solving gapped
  {H}amiltonians locally},\ }\bibfield  {journal} {\bibinfo  {journal}
  {Physical Review B}\ }\textbf {\bibinfo {volume} {73}},\ \href
  {https://doi.org/10.1103/physrevb.73.085115} {10.1103/physrevb.73.085115}
  (\bibinfo {year} {2006})\BibitemShut {NoStop}%
\bibitem [{\citenamefont {Verstraete}\ \emph {et~al.}(2004)\citenamefont
  {Verstraete}, \citenamefont {García-Ripoll},\ and\ \citenamefont
  {Cirac}}]{Verstraete_2004_mpo}%
  \BibitemOpen
  \bibfield  {author} {\bibinfo {author} {\bibfnamefont {F.}~\bibnamefont
  {Verstraete}}, \bibinfo {author} {\bibfnamefont {J.~J.}\ \bibnamefont
  {García-Ripoll}},\ and\ \bibinfo {author} {\bibfnamefont {J.~I.}\
  \bibnamefont {Cirac}},\ }\bibfield  {title} {\bibinfo {title} {Matrix product
  density operators: {S}imulation of finite-temperature and dissipative
  systems},\ }\bibfield  {journal} {\bibinfo  {journal} {Physical Review
  Letters}\ }\textbf {\bibinfo {volume} {93}},\ \href
  {https://doi.org/10.1103/physrevlett.93.207204}
  {10.1103/physrevlett.93.207204} (\bibinfo {year} {2004})\BibitemShut
  {NoStop}%
\bibitem [{\citenamefont {Cuevas}\ \emph {et~al.}(2013)\citenamefont {Cuevas},
  \citenamefont {Schuch}, \citenamefont {Pérez-García},\ and\ \citenamefont
  {Ignacio~Cirac}}]{Cuevas_2013_purification}%
  \BibitemOpen
  \bibfield  {author} {\bibinfo {author} {\bibfnamefont {G.~D.~l.}\
  \bibnamefont {Cuevas}}, \bibinfo {author} {\bibfnamefont {N.}~\bibnamefont
  {Schuch}}, \bibinfo {author} {\bibfnamefont {D.}~\bibnamefont
  {Pérez-García}},\ and\ \bibinfo {author} {\bibfnamefont {J.}~\bibnamefont
  {Ignacio~Cirac}},\ }\bibfield  {title} {\bibinfo {title} {Purifications of
  multipartite states: {L}imitations and constructive methods},\ }\href
  {https://doi.org/10.1088/1367-2630/15/12/123021} {\bibfield  {journal}
  {\bibinfo  {journal} {New Journal of Physics}\ }\textbf {\bibinfo {volume}
  {15}},\ \bibinfo {pages} {123021} (\bibinfo {year} {2013})}\BibitemShut
  {NoStop}%
\bibitem [{\citenamefont {Werner}\ \emph {et~al.}(2016)\citenamefont {Werner},
  \citenamefont {Jaschke}, \citenamefont {Silvi}, \citenamefont {Kliesch},
  \citenamefont {Calarco}, \citenamefont {Eisert},\ and\ \citenamefont
  {Montangero}}]{Werner_2016}%
  \BibitemOpen
  \bibfield  {author} {\bibinfo {author} {\bibfnamefont {A.}~\bibnamefont
  {Werner}}, \bibinfo {author} {\bibfnamefont {D.}~\bibnamefont {Jaschke}},
  \bibinfo {author} {\bibfnamefont {P.}~\bibnamefont {Silvi}}, \bibinfo
  {author} {\bibfnamefont {M.}~\bibnamefont {Kliesch}}, \bibinfo {author}
  {\bibfnamefont {T.}~\bibnamefont {Calarco}}, \bibinfo {author} {\bibfnamefont
  {J.}~\bibnamefont {Eisert}},\ and\ \bibinfo {author} {\bibfnamefont
  {S.}~\bibnamefont {Montangero}},\ }\bibfield  {title} {\bibinfo {title}
  {Positive tensor network approach for simulating open quantum many-body
  systems},\ }\bibfield  {journal} {\bibinfo  {journal} {Physical Review
  Letters}\ }\textbf {\bibinfo {volume} {116}},\ \href
  {https://doi.org/10.1103/physrevlett.116.237201}
  {10.1103/physrevlett.116.237201} (\bibinfo {year} {2016})\BibitemShut
  {NoStop}%
\bibitem [{\citenamefont {Suzuki}(1991)}]{trotter-higher-splitting}%
  \BibitemOpen
  \bibfield  {author} {\bibinfo {author} {\bibfnamefont {M.}~\bibnamefont
  {Suzuki}},\ }\bibfield  {title} {\bibinfo {title} {General theory of fractal
  path integrals with applications to many‐body theories and statistical
  physics},\ }\href {https://doi.org/10.1063/1.529425} {\bibfield  {journal}
  {\bibinfo  {journal} {Journal of Mathematical Physics}\ }\textbf {\bibinfo
  {volume} {32}},\ \bibinfo {pages} {400} (\bibinfo {year} {1991})},\ \Eprint
  {https://arxiv.org/abs/https://pubs.aip.org/aip/jmp/article-pdf/32/2/400/19166143/400\_1\_online.pdf}
  {https://pubs.aip.org/aip/jmp/article-pdf/32/2/400/19166143/400\_1\_online.pdf}
  \BibitemShut {NoStop}%
\bibitem [{\citenamefont {Absil}\ \emph {et~al.}(2008)\citenamefont {Absil},
  \citenamefont {Mahony},\ and\ \citenamefont
  {Sepulchre}}]{princeton_manifolds}%
  \BibitemOpen
  \bibfield  {author} {\bibinfo {author} {\bibfnamefont {P.-A.}\ \bibnamefont
  {Absil}}, \bibinfo {author} {\bibfnamefont {R.}~\bibnamefont {Mahony}},\ and\
  \bibinfo {author} {\bibfnamefont {R.}~\bibnamefont {Sepulchre}},\ }\href
  {http://www.jstor.org/stable/j.ctt7smmk} {\emph {\bibinfo {title}
  {Optimization Algorithms on Matrix Manifolds}}}\ (\bibinfo  {publisher}
  {Princeton University Press},\ \bibinfo {year} {2008})\BibitemShut {NoStop}%
\bibitem [{\citenamefont {Brieger}\ \emph {et~al.}(2023)\citenamefont
  {Brieger}, \citenamefont {Roth},\ and\ \citenamefont
  {Kliesch}}]{brieger_gst}%
  \BibitemOpen
  \bibfield  {author} {\bibinfo {author} {\bibfnamefont {R.}~\bibnamefont
  {Brieger}}, \bibinfo {author} {\bibfnamefont {I.}~\bibnamefont {Roth}},\ and\
  \bibinfo {author} {\bibfnamefont {M.}~\bibnamefont {Kliesch}},\ }\bibfield
  {title} {\bibinfo {title} {Compressive gate set tomography},\ }\href
  {https://doi.org/10.1103/PRXQuantum.4.010325} {\bibfield  {journal} {\bibinfo
   {journal} {PRX Quantum}\ }\textbf {\bibinfo {volume} {4}},\ \bibinfo {pages}
  {010325} (\bibinfo {year} {2023})}\BibitemShut {NoStop}%
\bibitem [{\citenamefont {Diehl}\ \emph {et~al.}(2011)\citenamefont {Diehl},
  \citenamefont {Rico}, \citenamefont {Baranov},\ and\ \citenamefont
  {Zoller}}]{kitaev_wire}%
  \BibitemOpen
  \bibfield  {author} {\bibinfo {author} {\bibfnamefont {S.}~\bibnamefont
  {Diehl}}, \bibinfo {author} {\bibfnamefont {E.}~\bibnamefont {Rico}},
  \bibinfo {author} {\bibfnamefont {M.~A.}\ \bibnamefont {Baranov}},\ and\
  \bibinfo {author} {\bibfnamefont {P.}~\bibnamefont {Zoller}},\ }\bibfield
  {title} {\bibinfo {title} {Topology by dissipation in atomic quantum wires},\
  }\href {https://doi.org/10.1038/nphys2106} {\bibfield  {journal} {\bibinfo
  {journal} {Nature Physics}\ }\textbf {\bibinfo {volume} {7}},\ \bibinfo
  {pages} {971–977} (\bibinfo {year} {2011})}\BibitemShut {NoStop}%
\bibitem [{\citenamefont {van~den Berg}\ \emph {et~al.}(2023)\citenamefont
  {van~den Berg}, \citenamefont {Minev}, \citenamefont {Kandala},\ and\
  \citenamefont {Temme}}]{pec_spl}%
  \BibitemOpen
  \bibfield  {author} {\bibinfo {author} {\bibfnamefont {E.}~\bibnamefont
  {van~den Berg}}, \bibinfo {author} {\bibfnamefont {Z.~K.}\ \bibnamefont
  {Minev}}, \bibinfo {author} {\bibfnamefont {A.}~\bibnamefont {Kandala}},\
  and\ \bibinfo {author} {\bibfnamefont {K.}~\bibnamefont {Temme}},\ }\bibfield
   {title} {\bibinfo {title} {Probabilistic error cancellation with sparse
  {Pauli–Lindblad} models on noisy quantum processors},\ }\href
  {https://doi.org/10.1038/s41567-023-02042-2} {\bibfield  {journal} {\bibinfo
  {journal} {Nature Physics}\ }\textbf {\bibinfo {volume} {19}},\ \bibinfo
  {pages} {1116–1121} (\bibinfo {year} {2023})}\BibitemShut {NoStop}%
\bibitem [{\citenamefont {Fannes}\ \emph {et~al.}(1992)\citenamefont {Fannes},
  \citenamefont {Nachtergaele},\ and\ \citenamefont
  {Werner}}]{Fannes_Nachtergaele_Werner_1992_mps}%
  \BibitemOpen
  \bibfield  {author} {\bibinfo {author} {\bibfnamefont {M.}~\bibnamefont
  {Fannes}}, \bibinfo {author} {\bibfnamefont {B.}~\bibnamefont
  {Nachtergaele}},\ and\ \bibinfo {author} {\bibfnamefont {R.~F.}\ \bibnamefont
  {Werner}},\ }\bibfield  {title} {\bibinfo {title} {Finitely correlated states
  on quantum spin chains},\ }\href {https://doi.org/10.1007/bf02099178}
  {\bibfield  {journal} {\bibinfo  {journal} {Communications in Mathematical
  Physics}\ }\textbf {\bibinfo {volume} {144}},\ \bibinfo {pages} {443–490}
  (\bibinfo {year} {1992})}\BibitemShut {NoStop}%
\bibitem [{\citenamefont {Perez-Garcia}\ \emph {et~al.}(2007)\citenamefont
  {Perez-Garcia}, \citenamefont {Verstraete}, \citenamefont {Wolf},\ and\
  \citenamefont {Cirac}}]{perezgarcia2007matrix}%
  \BibitemOpen
  \bibfield  {author} {\bibinfo {author} {\bibfnamefont {D.}~\bibnamefont
  {Perez-Garcia}}, \bibinfo {author} {\bibfnamefont {F.}~\bibnamefont
  {Verstraete}}, \bibinfo {author} {\bibfnamefont {M.~M.}\ \bibnamefont
  {Wolf}},\ and\ \bibinfo {author} {\bibfnamefont {J.~I.}\ \bibnamefont
  {Cirac}},\ }\href@noop {} {\bibinfo {title} {Matrix product state
  representations}} (\bibinfo {year} {2007}),\ \Eprint
  {https://arxiv.org/abs/quant-ph/0608197} {arXiv:quant-ph/0608197 [quant-ph]}
  \BibitemShut {NoStop}%
\bibitem [{\citenamefont {Zwolak}\ and\ \citenamefont
  {Vidal}(2004)}]{Zwolak_2004_mpo}%
  \BibitemOpen
  \bibfield  {author} {\bibinfo {author} {\bibfnamefont {M.}~\bibnamefont
  {Zwolak}}\ and\ \bibinfo {author} {\bibfnamefont {G.}~\bibnamefont {Vidal}},\
  }\bibfield  {title} {\bibinfo {title} {Mixed-state dynamics in
  one-dimensional quantum lattice systems: {A} time-dependent superoperator
  renormalization algorithm},\ }\bibfield  {journal} {\bibinfo  {journal}
  {Physical Review Letters}\ }\textbf {\bibinfo {volume} {93}},\ \href
  {https://doi.org/10.1103/physrevlett.93.207205}
  {10.1103/physrevlett.93.207205} (\bibinfo {year} {2004})\BibitemShut
  {NoStop}%
\bibitem [{\citenamefont {Kliesch}\ \emph {et~al.}(2014)\citenamefont
  {Kliesch}, \citenamefont {Gross},\ and\ \citenamefont
  {Eisert}}]{Kliesch_positive_np}%
  \BibitemOpen
  \bibfield  {author} {\bibinfo {author} {\bibfnamefont {M.}~\bibnamefont
  {Kliesch}}, \bibinfo {author} {\bibfnamefont {D.}~\bibnamefont {Gross}},\
  and\ \bibinfo {author} {\bibfnamefont {J.}~\bibnamefont {Eisert}},\
  }\bibfield  {title} {\bibinfo {title} {Matrix-product operators and states:
  {NP}-hardness and undecidability},\ }\href
  {https://doi.org/10.1103/PhysRevLett.113.160503} {\bibfield  {journal}
  {\bibinfo  {journal} {Phys. Rev. Lett.}\ }\textbf {\bibinfo {volume} {113}},\
  \bibinfo {pages} {160503} (\bibinfo {year} {2014})}\BibitemShut {NoStop}%
\bibitem [{\citenamefont {Pichler}\ \emph {et~al.}(2013)\citenamefont
  {Pichler}, \citenamefont {Schachenmayer}, \citenamefont {Daley},\ and\
  \citenamefont {Zoller}}]{optics_trajectories}%
  \BibitemOpen
  \bibfield  {author} {\bibinfo {author} {\bibfnamefont {H.}~\bibnamefont
  {Pichler}}, \bibinfo {author} {\bibfnamefont {J.}~\bibnamefont
  {Schachenmayer}}, \bibinfo {author} {\bibfnamefont {A.~J.}\ \bibnamefont
  {Daley}},\ and\ \bibinfo {author} {\bibfnamefont {P.}~\bibnamefont
  {Zoller}},\ }\bibfield  {title} {\bibinfo {title} {Heating dynamics of
  bosonic atoms in a noisy optical lattice},\ }\href
  {https://doi.org/10.1103/PhysRevA.87.033606} {\bibfield  {journal} {\bibinfo
  {journal} {Phys. Rev. A}\ }\textbf {\bibinfo {volume} {87}},\ \bibinfo
  {pages} {033606} (\bibinfo {year} {2013})}\BibitemShut {NoStop}%
\bibitem [{\citenamefont {Vidal}(2003)}]{Vidal_2003_tebd}%
  \BibitemOpen
  \bibfield  {author} {\bibinfo {author} {\bibfnamefont {G.}~\bibnamefont
  {Vidal}},\ }\bibfield  {title} {\bibinfo {title} {Efficient classical
  simulation of slightly entangled quantum computations},\ }\bibfield
  {journal} {\bibinfo  {journal} {Physical Review Letters}\ }\textbf {\bibinfo
  {volume} {91}},\ \href {https://doi.org/10.1103/physrevlett.91.147902}
  {10.1103/physrevlett.91.147902} (\bibinfo {year} {2003})\BibitemShut
  {NoStop}%
\bibitem [{\citenamefont {Kliesch}\ \emph {et~al.}(2011)\citenamefont
  {Kliesch}, \citenamefont {Barthel}, \citenamefont {Gogolin}, \citenamefont
  {Kastoryano},\ and\ \citenamefont {Eisert}}]{kliesh_trotter_open}%
  \BibitemOpen
  \bibfield  {author} {\bibinfo {author} {\bibfnamefont {M.}~\bibnamefont
  {Kliesch}}, \bibinfo {author} {\bibfnamefont {T.}~\bibnamefont {Barthel}},
  \bibinfo {author} {\bibfnamefont {C.}~\bibnamefont {Gogolin}}, \bibinfo
  {author} {\bibfnamefont {M.}~\bibnamefont {Kastoryano}},\ and\ \bibinfo
  {author} {\bibfnamefont {J.}~\bibnamefont {Eisert}},\ }\bibfield  {title}
  {\bibinfo {title} {Dissipative quantum {C}hurch-{T}uring theorem},\ }\href
  {https://doi.org/10.1103/PhysRevLett.107.120501} {\bibfield  {journal}
  {\bibinfo  {journal} {Phys. Rev. Lett.}\ }\textbf {\bibinfo {volume} {107}},\
  \bibinfo {pages} {120501} (\bibinfo {year} {2011})}\BibitemShut {NoStop}%
\bibitem [{\citenamefont {Wood}\ \emph {et~al.}(2015)\citenamefont {Wood},
  \citenamefont {Biamonte},\ and\ \citenamefont {Cory}}]{wood2015tensor}%
  \BibitemOpen
  \bibfield  {author} {\bibinfo {author} {\bibfnamefont {C.~J.}\ \bibnamefont
  {Wood}}, \bibinfo {author} {\bibfnamefont {J.~D.}\ \bibnamefont {Biamonte}},\
  and\ \bibinfo {author} {\bibfnamefont {D.~G.}\ \bibnamefont {Cory}},\
  }\href@noop {} {\bibinfo {title} {Tensor networks and graphical calculus for
  open quantum systems}} (\bibinfo {year} {2015}),\ \Eprint
  {https://arxiv.org/abs/1111.6950} {arXiv:1111.6950 [quant-ph]} \BibitemShut
  {NoStop}%
\bibitem [{\citenamefont {Edelman}\ \emph {et~al.}(1998)\citenamefont
  {Edelman}, \citenamefont {Arias},\ and\ \citenamefont
  {Smith}}]{edelman1998geometry_stiefel}%
  \BibitemOpen
  \bibfield  {author} {\bibinfo {author} {\bibfnamefont {A.}~\bibnamefont
  {Edelman}}, \bibinfo {author} {\bibfnamefont {T.~A.}\ \bibnamefont {Arias}},\
  and\ \bibinfo {author} {\bibfnamefont {S.~T.}\ \bibnamefont {Smith}},\
  }\href@noop {} {\bibinfo {title} {The geometry of algorithms with
  orthogonality constraints}} (\bibinfo {year} {1998}),\ \Eprint
  {https://arxiv.org/abs/physics/9806030} {arXiv:physics/9806030
  [physics.comp-ph]} \BibitemShut {NoStop}%
\bibitem [{\citenamefont {Nguyen}(2021)}]{nguyenoperatorvalued_metric_based}%
  \BibitemOpen
  \bibfield  {author} {\bibinfo {author} {\bibfnamefont {D.}~\bibnamefont
  {Nguyen}},\ }\href@noop {} {\bibinfo {title} {Operator-valued formulas for
  {R}iemannian gradient and {H}essian and families of tractable metrics}}
  (\bibinfo {year} {2021}),\ \Eprint {https://arxiv.org/abs/2009.10159}
  {arXiv:2009.10159 [math.OC]} \BibitemShut {NoStop}%
\bibitem [{\citenamefont {Zhang}(2020)}]{zhang2020newton_retraction}%
  \BibitemOpen
  \bibfield  {author} {\bibinfo {author} {\bibfnamefont {R.}~\bibnamefont
  {Zhang}},\ }\href@noop {} {\bibinfo {title} {Newton retraction as approximate
  geodesics on submanifolds}} (\bibinfo {year} {2020}),\ \Eprint
  {https://arxiv.org/abs/2006.14751} {arXiv:2006.14751 [math.NA]} \BibitemShut
  {NoStop}%
\bibitem [{\citenamefont {Yoshida}(1990)}]{YOSHIDA1990262}%
  \BibitemOpen
  \bibfield  {author} {\bibinfo {author} {\bibfnamefont {H.}~\bibnamefont
  {Yoshida}},\ }\bibfield  {title} {\bibinfo {title} {Construction of higher
  order symplectic integrators},\ }\href
  {https://doi.org/https://doi.org/10.1016/0375-9601(90)90092-3} {\bibfield
  {journal} {\bibinfo  {journal} {Physics Letters A}\ }\textbf {\bibinfo
  {volume} {150}},\ \bibinfo {pages} {262} (\bibinfo {year}
  {1990})}\BibitemShut {NoStop}%
\bibitem [{\citenamefont {Blanes}\ and\ \citenamefont
  {Moan}(2002)}]{BLANES2002313}%
  \BibitemOpen
  \bibfield  {author} {\bibinfo {author} {\bibfnamefont {S.}~\bibnamefont
  {Blanes}}\ and\ \bibinfo {author} {\bibfnamefont {P.}~\bibnamefont {Moan}},\
  }\bibfield  {title} {\bibinfo {title} {Practical symplectic partitioned
  {Runge–Kutta} and {Runge–Kutta–Nyström} methods},\ }\href
  {https://doi.org/https://doi.org/10.1016/S0377-0427(01)00492-7} {\bibfield
  {journal} {\bibinfo  {journal} {Journal of Computational and Applied
  Mathematics}\ }\textbf {\bibinfo {volume} {142}},\ \bibinfo {pages} {313}
  (\bibinfo {year} {2002})}\BibitemShut {NoStop}%
\bibitem [{\citenamefont {Luchnikov}\ \emph {et~al.}(2021)\citenamefont
  {Luchnikov}, \citenamefont {Ryzhov}, \citenamefont {Filippov},\ and\
  \citenamefont {Ouerdane}}]{Luchnikov_2021}%
  \BibitemOpen
  \bibfield  {author} {\bibinfo {author} {\bibfnamefont {I.}~\bibnamefont
  {Luchnikov}}, \bibinfo {author} {\bibfnamefont {A.}~\bibnamefont {Ryzhov}},
  \bibinfo {author} {\bibfnamefont {S.}~\bibnamefont {Filippov}},\ and\
  \bibinfo {author} {\bibfnamefont {H.}~\bibnamefont {Ouerdane}},\ }\bibfield
  {title} {\bibinfo {title} {Qgopt: Riemannian optimization for quantum
  technologies},\ }\bibfield  {journal} {\bibinfo  {journal} {SciPost Physics}\
  }\textbf {\bibinfo {volume} {10}},\ \href
  {https://doi.org/10.21468/scipostphys.10.3.079}
  {10.21468/scipostphys.10.3.079} (\bibinfo {year} {2021})\BibitemShut
  {NoStop}%
\bibitem [{\citenamefont {{Python Software Foundation}}(2022)}]{python311}%
  \BibitemOpen
  \bibfield  {author} {\bibinfo {author} {\bibnamefont {{Python Software
  Foundation}}},\ }\href {https://www.python.org/downloads/release/python-311/}
  {\bibinfo {title} {{Python 3.11}}},\ \bibinfo {howpublished} {[Computer
  software]} (\bibinfo {year} {2022})\BibitemShut {NoStop}%
\bibitem [{\citenamefont {Godinez}(2023)}]{opentn}%
  \BibitemOpen
  \bibfield  {author} {\bibinfo {author} {\bibfnamefont {E.}~\bibnamefont
  {Godinez}},\ }\href {https://github.com/EmilianoG-byte/opentn} {\bibinfo
  {title} {{OpenTN}: {A} riemannian approach to the {L}indbladian dynamics of a
  locally purified tensor network}},\ \bibinfo {howpublished} {[Online]}
  (\bibinfo {year} {2023})\BibitemShut {NoStop}%
\bibitem [{\citenamefont {Harris}\ \emph {et~al.}(2020)\citenamefont {Harris},
  \citenamefont {Millman}, \citenamefont {van~der Walt}, \citenamefont
  {Gommers}, \citenamefont {Virtanen}, \citenamefont {Cournapeau},
  \citenamefont {Wieser}, \citenamefont {Taylor}, \citenamefont {Berg},
  \citenamefont {Smith}, \citenamefont {Kern}, \citenamefont {Picus},
  \citenamefont {Hoyer}, \citenamefont {van Kerkwijk}, \citenamefont {Brett},
  \citenamefont {Haldane}, \citenamefont {del R{\'{i}}o}, \citenamefont
  {Wiebe}, \citenamefont {Peterson}, \citenamefont {G{\'{e}}rard-Marchant},
  \citenamefont {Sheppard}, \citenamefont {Reddy}, \citenamefont {Weckesser},
  \citenamefont {Abbasi}, \citenamefont {Gohlke},\ and\ \citenamefont
  {Oliphant}}]{harris2020array}%
  \BibitemOpen
  \bibfield  {author} {\bibinfo {author} {\bibfnamefont {C.~R.}\ \bibnamefont
  {Harris}}, \bibinfo {author} {\bibfnamefont {K.~J.}\ \bibnamefont {Millman}},
  \bibinfo {author} {\bibfnamefont {S.~J.}\ \bibnamefont {van~der Walt}},
  \bibinfo {author} {\bibfnamefont {R.}~\bibnamefont {Gommers}}, \bibinfo
  {author} {\bibfnamefont {P.}~\bibnamefont {Virtanen}}, \bibinfo {author}
  {\bibfnamefont {D.}~\bibnamefont {Cournapeau}}, \bibinfo {author}
  {\bibfnamefont {E.}~\bibnamefont {Wieser}}, \bibinfo {author} {\bibfnamefont
  {J.}~\bibnamefont {Taylor}}, \bibinfo {author} {\bibfnamefont
  {S.}~\bibnamefont {Berg}}, \bibinfo {author} {\bibfnamefont {N.~J.}\
  \bibnamefont {Smith}}, \bibinfo {author} {\bibfnamefont {R.}~\bibnamefont
  {Kern}}, \bibinfo {author} {\bibfnamefont {M.}~\bibnamefont {Picus}},
  \bibinfo {author} {\bibfnamefont {S.}~\bibnamefont {Hoyer}}, \bibinfo
  {author} {\bibfnamefont {M.~H.}\ \bibnamefont {van Kerkwijk}}, \bibinfo
  {author} {\bibfnamefont {M.}~\bibnamefont {Brett}}, \bibinfo {author}
  {\bibfnamefont {A.}~\bibnamefont {Haldane}}, \bibinfo {author} {\bibfnamefont
  {J.~F.}\ \bibnamefont {del R{\'{i}}o}}, \bibinfo {author} {\bibfnamefont
  {M.}~\bibnamefont {Wiebe}}, \bibinfo {author} {\bibfnamefont
  {P.}~\bibnamefont {Peterson}}, \bibinfo {author} {\bibfnamefont
  {P.}~\bibnamefont {G{\'{e}}rard-Marchant}}, \bibinfo {author} {\bibfnamefont
  {K.}~\bibnamefont {Sheppard}}, \bibinfo {author} {\bibfnamefont
  {T.}~\bibnamefont {Reddy}}, \bibinfo {author} {\bibfnamefont
  {W.}~\bibnamefont {Weckesser}}, \bibinfo {author} {\bibfnamefont
  {H.}~\bibnamefont {Abbasi}}, \bibinfo {author} {\bibfnamefont
  {C.}~\bibnamefont {Gohlke}},\ and\ \bibinfo {author} {\bibfnamefont {T.~E.}\
  \bibnamefont {Oliphant}},\ }\bibfield  {title} {\bibinfo {title} {Array
  programming with {NumPy}},\ }\href
  {https://doi.org/10.1038/s41586-020-2649-2} {\bibfield  {journal} {\bibinfo
  {journal} {Nature}\ }\textbf {\bibinfo {volume} {585}},\ \bibinfo {pages}
  {357} (\bibinfo {year} {2020})}\BibitemShut {NoStop}%
\bibitem [{\citenamefont {Bradbury}\ \emph {et~al.}(2018)\citenamefont
  {Bradbury}, \citenamefont {Frostig}, \citenamefont {Hawkins}, \citenamefont
  {Johnson}, \citenamefont {Leary}, \citenamefont {Maclaurin}, \citenamefont
  {Necula}, \citenamefont {Paszke}, \citenamefont {Vander{P}las}, \citenamefont
  {Wanderman-{M}ilne},\ and\ \citenamefont {Zhang}}]{jax2018github}%
  \BibitemOpen
  \bibfield  {author} {\bibinfo {author} {\bibfnamefont {J.}~\bibnamefont
  {Bradbury}}, \bibinfo {author} {\bibfnamefont {R.}~\bibnamefont {Frostig}},
  \bibinfo {author} {\bibfnamefont {P.}~\bibnamefont {Hawkins}}, \bibinfo
  {author} {\bibfnamefont {M.~J.}\ \bibnamefont {Johnson}}, \bibinfo {author}
  {\bibfnamefont {C.}~\bibnamefont {Leary}}, \bibinfo {author} {\bibfnamefont
  {D.}~\bibnamefont {Maclaurin}}, \bibinfo {author} {\bibfnamefont
  {G.}~\bibnamefont {Necula}}, \bibinfo {author} {\bibfnamefont
  {A.}~\bibnamefont {Paszke}}, \bibinfo {author} {\bibfnamefont
  {J.}~\bibnamefont {Vander{P}las}}, \bibinfo {author} {\bibfnamefont
  {S.}~\bibnamefont {Wanderman-{M}ilne}},\ and\ \bibinfo {author}
  {\bibfnamefont {Q.}~\bibnamefont {Zhang}},\ }\href
  {http://github.com/google/jax} {\bibinfo {title} {{JAX}: composable
  transformations of {P}ython+{N}um{P}y programs}} (\bibinfo {year}
  {2018})\BibitemShut {NoStop}%
\bibitem [{\citenamefont {Manton}(2002)}]{unitary_constraints}%
  \BibitemOpen
  \bibfield  {author} {\bibinfo {author} {\bibfnamefont {J.}~\bibnamefont
  {Manton}},\ }\bibfield  {title} {\bibinfo {title} {Optimization algorithms
  exploiting unitary constraints},\ }\href {https://doi.org/10.1109/78.984753}
  {\bibfield  {journal} {\bibinfo  {journal} {IEEE Transactions on Signal
  Processing}\ }\textbf {\bibinfo {volume} {50}},\ \bibinfo {pages} {635}
  (\bibinfo {year} {2002})}\BibitemShut {NoStop}%
\bibitem [{\citenamefont {Cao}\ and\ \citenamefont
  {Lu}(2021)}]{cao2021structurepreserving}%
  \BibitemOpen
  \bibfield  {author} {\bibinfo {author} {\bibfnamefont {Y.}~\bibnamefont
  {Cao}}\ and\ \bibinfo {author} {\bibfnamefont {J.}~\bibnamefont {Lu}},\
  }\href@noop {} {\bibinfo {title} {Structure-preserving numerical schemes for
  {L}indblad equations}} (\bibinfo {year} {2021}),\ \Eprint
  {https://arxiv.org/abs/2103.01194} {arXiv:2103.01194 [math.NA]} \BibitemShut
  {NoStop}%
\bibitem [{\citenamefont {Cheng}\ \emph {et~al.}(2021)\citenamefont {Cheng},
  \citenamefont {Cao}, \citenamefont {Zhang}, \citenamefont {Liu},
  \citenamefont {Hou}, \citenamefont {Xu},\ and\ \citenamefont
  {Zeng}}]{Cheng_2021}%
  \BibitemOpen
  \bibfield  {author} {\bibinfo {author} {\bibfnamefont {S.}~\bibnamefont
  {Cheng}}, \bibinfo {author} {\bibfnamefont {C.}~\bibnamefont {Cao}}, \bibinfo
  {author} {\bibfnamefont {C.}~\bibnamefont {Zhang}}, \bibinfo {author}
  {\bibfnamefont {Y.}~\bibnamefont {Liu}}, \bibinfo {author} {\bibfnamefont
  {S.-Y.}\ \bibnamefont {Hou}}, \bibinfo {author} {\bibfnamefont
  {P.}~\bibnamefont {Xu}},\ and\ \bibinfo {author} {\bibfnamefont
  {B.}~\bibnamefont {Zeng}},\ }\bibfield  {title} {\bibinfo {title} {Simulating
  noisy quantum circuits with matrix product density operators},\ }\bibfield
  {journal} {\bibinfo  {journal} {Physical Review Research}\ }\textbf {\bibinfo
  {volume} {3}},\ \href {https://doi.org/10.1103/physrevresearch.3.023005}
  {10.1103/physrevresearch.3.023005} (\bibinfo {year} {2021})\BibitemShut
  {NoStop}%
\bibitem [{\citenamefont {Müller}\ \emph {et~al.}(2024)\citenamefont
  {Müller}, \citenamefont {Ayral},\ and\ \citenamefont
  {Bertrand}}]{müller2024enabling}%
  \BibitemOpen
  \bibfield  {author} {\bibinfo {author} {\bibfnamefont {A.}~\bibnamefont
  {Müller}}, \bibinfo {author} {\bibfnamefont {T.}~\bibnamefont {Ayral}},\
  and\ \bibinfo {author} {\bibfnamefont {C.}~\bibnamefont {Bertrand}},\
  }\href@noop {} {\bibinfo {title} {Enabling large-depth simulation of noisy
  quantum circuits with positive tensor networks}} (\bibinfo {year} {2024}),\
  \Eprint {https://arxiv.org/abs/2403.00152} {arXiv:2403.00152 [quant-ph]}
  \BibitemShut {NoStop}%
\bibitem [{\citenamefont {Kotil}\ \emph {et~al.}(2023)\citenamefont {Kotil},
  \citenamefont {Banerjee}, \citenamefont {Huang},\ and\ \citenamefont
  {Mendl}}]{kotil2023riemannian}%
  \BibitemOpen
  \bibfield  {author} {\bibinfo {author} {\bibfnamefont {A.}~\bibnamefont
  {Kotil}}, \bibinfo {author} {\bibfnamefont {R.}~\bibnamefont {Banerjee}},
  \bibinfo {author} {\bibfnamefont {Q.}~\bibnamefont {Huang}},\ and\ \bibinfo
  {author} {\bibfnamefont {C.~B.}\ \bibnamefont {Mendl}},\ }\href@noop {}
  {\bibinfo {title} {Riemannian quantum circuit optimization for {H}amiltonian
  simulation}} (\bibinfo {year} {2023}),\ \Eprint
  {https://arxiv.org/abs/2212.07556} {arXiv:2212.07556 [quant-ph]} \BibitemShut
  {NoStop}%
\bibitem [{\citenamefont {Dereniowski}\ and\ \citenamefont
  {Kubale}(2004)}]{cholesky}%
  \BibitemOpen
  \bibfield  {author} {\bibinfo {author} {\bibfnamefont {D.}~\bibnamefont
  {Dereniowski}}\ and\ \bibinfo {author} {\bibfnamefont {M.}~\bibnamefont
  {Kubale}},\ }\bibfield  {title} {\bibinfo {title} {Cholesky factorization of
  matrices in parallel and ranking of graphs},\ }in\ \href@noop {} {\emph
  {\bibinfo {booktitle} {Parallel Processing and Applied Mathematics}}},\
  \bibinfo {editor} {edited by\ \bibinfo {editor} {\bibfnamefont
  {R.}~\bibnamefont {Wyrzykowski}}, \bibinfo {editor} {\bibfnamefont
  {J.}~\bibnamefont {Dongarra}}, \bibinfo {editor} {\bibfnamefont
  {M.}~\bibnamefont {Paprzycki}},\ and\ \bibinfo {editor} {\bibfnamefont
  {J.}~\bibnamefont {Wa{\'{s}}niewski}}}\ (\bibinfo  {publisher} {Springer
  Berlin Heidelberg},\ \bibinfo {address} {Berlin, Heidelberg},\ \bibinfo
  {year} {2004})\ pp.\ \bibinfo {pages} {985--992}\BibitemShut {NoStop}%
\bibitem [{\citenamefont {Hauru}\ \emph {et~al.}(2021)\citenamefont {Hauru},
  \citenamefont {Van~Damme},\ and\ \citenamefont {Haegeman}}]{Hauru_2021}%
  \BibitemOpen
  \bibfield  {author} {\bibinfo {author} {\bibfnamefont {M.}~\bibnamefont
  {Hauru}}, \bibinfo {author} {\bibfnamefont {M.}~\bibnamefont {Van~Damme}},\
  and\ \bibinfo {author} {\bibfnamefont {J.}~\bibnamefont {Haegeman}},\
  }\bibfield  {title} {\bibinfo {title} {Riemannian optimization of isometric
  tensor networks},\ }\bibfield  {journal} {\bibinfo  {journal} {SciPost
  Physics}\ }\textbf {\bibinfo {volume} {10}},\ \href
  {https://doi.org/10.21468/scipostphys.10.2.040}
  {10.21468/scipostphys.10.2.040} (\bibinfo {year} {2021})\BibitemShut
  {NoStop}%
\bibitem [{\citenamefont {Higham}\ \emph {et~al.}(2010)\citenamefont {Higham},
  \citenamefont {Mehl},\ and\ \citenamefont {Tisseur}}]{canonical-polar}%
  \BibitemOpen
  \bibfield  {author} {\bibinfo {author} {\bibfnamefont {N.~J.}\ \bibnamefont
  {Higham}}, \bibinfo {author} {\bibfnamefont {C.}~\bibnamefont {Mehl}},\ and\
  \bibinfo {author} {\bibfnamefont {F.}~\bibnamefont {Tisseur}},\ }\bibfield
  {title} {\bibinfo {title} {The canonical generalized polar decomposition},\
  }\href {https://doi.org/10.1137/090765018} {\bibfield  {journal} {\bibinfo
  {journal} {SIAM Journal on Matrix Analysis and Applications}\ }\textbf
  {\bibinfo {volume} {31}},\ \bibinfo {pages} {2163} (\bibinfo {year}
  {2010})},\ \Eprint {https://arxiv.org/abs/https://doi.org/10.1137/090765018}
  {https://doi.org/10.1137/090765018} \BibitemShut {NoStop}%
\end{thebibliography}%

\appendix
\section{\label{app:transformations} Quantum channel transformations}
In this appendix, we follow a similar set of conventions as the ones used in \cite{wood2015tensor} with the following definitions:

\begin{definition}[Hilbert spaces]\label{def:hilbert-spaces}
    $\{\H_i\}_{i\in \N}$ are finite dimensional real Hilbert spaces of dimension $d_i$. We will omit the index $i$ whenever we can infer the Hilbert space from the context without ambiguity.
\end{definition}

\begin{definition}[Linear operators]\label{def:linear-operators}
    $\fancyL(\H_1, \H_2)$ is the space of bounded linear operators from $\H_1$ to $\H_2$
    \begin{equation}
        \label{eq:linear-operators}
        \Lambda: \H_1 \rightarrow \H_2
    \end{equation}
    with $\fancyL(\H_1) := \fancyL(\H_1, \H_1)$. Writing $\Lambda \in \fancyL(\H_1, \H_2)$ means that $\Lambda$ is a $d_1 \times d_2$ matrix.
\end{definition}

\begin{definition}[Superoperators]\label{def:superoperators}
    Also known as operator maps, $\T(\H_1, \H_2)$ is the space of linear maps from operators to operators
    \begin{equation}
        \label{eq:superoperators}
        \Phi: \fancyL(\H_1) \rightarrow \fancyL(\H_2)
    \end{equation}
    with $\T(\H_1) := \T(\H_1, \H_1)$.
\end{definition}

\section{\label{app:form-global-local} From global to local superoperators}
Let us look in detail at the form of the odd $\e^{\tau \hat{\L_o}}$ and even $\e^{\tau \hat{\L_e}}$ layers in the splitting shown in \cref{eq:trotter_two_sites} for a quantum system with $N$ sites. 
Each superoperator $\hat{\L}^{[\ell,\ell+1]} \in \R^{d^{2N} \times d^{2N}}$ in \cref{eq:full_local_Liouvillian} is of the form described in \cref{eq:lindblad-vectorized},
\begin{multline} \label{eq:local_liouvillian}
    \hat{\L}^{[\ell,\ell+1]} = \sum_k \left[ L_k^{[\ell,\ell+1]} \otimes L_k^{[\ell,\ell+1] *} - \right. \\
    \left. \frac{1}{2}\left(L^{[\ell,\ell+1] \dagger}_k L_k^{[\ell,\ell+1]} \otimes \I^{\otimes N} + \I^{\otimes N} \otimes L^{[\ell,\ell+1] T}_k L^{[\ell,\ell+1] *}_k \right) \right] \, ,
\end{multline}
where  $\I^{\otimes N}$ is a $d^N \times d^N $ identity matrix. The jump operators $L_k^{[\ell,\ell+1]}$ in \cref{eq:local_liouvillian} are a special case of the ones we saw in \cref{eq:lindblad-vectorized}, as they are defined on a Hilbert space of dimension ${d^N}$ while acting non-trivially only on two sites at a time: $\ell$ and $\ell+1$. Namely, they are of the form
\begin{align}
    L_k^{[\ell, \ell+1]} = \I^{\otimes \ell-1} \otimes L_k \otimes \I^{\otimes N - 1 - \ell}, \qquad L_k \in \R^{d^2 \times d^2},\label{eq:lindblad-local-global}
\end{align}
for $\ell \in [1, N-1]$, while for $\ell = N$, we use periodic boundary conditions (PBC) to define
\begin{align}
    L_k^{[N, 1]} = L_k^{1} \otimes \I^{\otimes N-2} \otimes L_k^{N}, \quad L_k = L_k^N \otimes L_k^{1} \in \R^{d^2 \times d^2} \, .\label{eq:lindblad-local-global-pbc}
\end{align}

The diagram representation of the boundary operator showing the PBC is
\begin{equation}
    L_k^{[N, 1]} = \GraphicEquationH{0.5}{0.1}{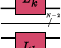} \, .
    \label{eq:lk_periodic}
\end{equation}

Using \cref{eq:local_liouvillian,eq:lindblad-local-global} we can rewrite each $ \hat{\L}^{[\ell,\ell+1]}$ as
\begin{multline} \label{eq:local_liouvillian_two_site}
    \hat{\L}^{[\ell,\ell+1]} = \sum_k \\
    \Big[ \left( \I^{\otimes \ell-1} \otimes L_k \otimes \I^{\otimes N - 1 - \ell} \right) \otimes \left( \I^{\otimes \ell-1} \otimes L_k^* \otimes \I^{\otimes N - 1 - \ell} \right)  \\
    - \left( \I^{\otimes \ell-1} \otimes L_k^\dagger L_k \otimes \I^{\otimes N - 1 - \ell}\right) \otimes \frac{\I^{\otimes N}}{2}  \\
    - \frac{\I^{\otimes N}}{2}  \otimes \left(\I^{\otimes \ell-1} \otimes L_k^T L_k^* \otimes \I^{\otimes N - 1 - \ell} \right) \Big] \,,
\end{multline}
and the tensor representation for the $k$-th operator is
\begin{widetext}
\begin{equation}
    \GraphicEquationH{0.47}{0.2075}{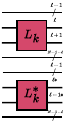} - \frac{1}{2} \left( \GraphicEquationH{0.45}{0.2}{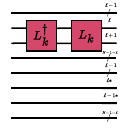} + \GraphicEquationH{0.45}{0.2}{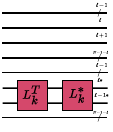} \right) \, .
    \label{eq:lindblad-full-vec-diagram}
\end{equation}
\end{widetext}
In this diagram, we recognize the non-identity terms as the vectorization of the $k$-th dissipative term $\hat{\L}_k^{[\ell, \ell+1]}$ in \cref{eq:lindblad-vectorized}. Contrary to the single-site scenario, the components of this operator in a multi-site system are spread apart (see global vectorization in \cref{eq:rho_n_sites_vec_global}). This difference invites us to define the tensors $\D^{k [\ell, \ell+1]}_{1, j}$ and $\D^{k [\ell, \ell+1]}_{2, j}$ as
\begin{equation}
    \begin{aligned}
    \D^{k [\ell, \ell+1]}_{1, 1} &= L_k \, , \\
    \D^{k [\ell, \ell+1]}_{1, 2} &= \frac{1}{\sqrt{2}} L_k^{\dagger} L_k \, , \\
    \D^{k [\ell, \ell+1]}_{1, 3} &= \frac{1}{\sqrt{2}} \I \, ,
    \end{aligned}
\end{equation}
and
\begin{equation}
    \begin{aligned}
    \D^{k [\ell, \ell+1]*}_{2, 1} &= L_k^* \, , \\
    \D^{k [\ell, \ell+1]*}_{2, 2} &= \frac{1}{\sqrt{2}} L_k^{T} L_k^* \, , \\
    \D^{k [\ell, \ell+1]*}_{2, 3} &= \frac{1}{\sqrt{2}} \I \, ,
    \end{aligned}
\end{equation}
respectively. Using these tensors, we arrive at the expression
\begin{equation}
    \hat{\D}_k^{[\ell, \ell+1]} = \sum_j \D^{k [\ell, \ell+1]}_{1, j} \otimes \D^{k [\ell, \ell+1]*}_{2, j}.
    \label{eq:dissipative-components}
\end{equation}
By summing over all $k$ we obtain
\begin{equation}
    \hat{\D}^{[\ell, \ell+1]} = \sum_k \D_k^{[\ell, \ell+1]} = \sum_{j, k} \D^{k [\ell, \ell+1]}_{1, j} \otimes \D^{k [\ell, \ell+1]*}_{2, j} \, ,
    \label{eq:dissipative-sum-two-sites}
\end{equation}
with the tensor diagram representation
\begin{equation}
    \GraphicEquationH{0.455}{0.1}{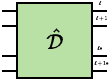} = \GraphicEquationH{0.455}{0.1}{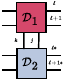} \, .
    \label{eq:dissipative-sum-two-sites-diagram}
\end{equation}
Using these relations, we can rewrite the global superoperator $\hat{\L}^{[\ell,\ell+1]}$ as a tensor product of the form
\begin{equation}
    \begin{aligned}
        \hat{\L}^{[\ell,\ell+1]} &=
        \sum_{j, k} \I^{\otimes \ell-1} \otimes \D^{k [\ell, \ell+1]}_{1, j} \otimes \I^{\otimes N - 1 - \ell} \\
        & \otimes \I^{\otimes \ell-1} \otimes \D^{k [\ell, \ell+1]*}_{2, j} \otimes \I^{\otimes N - 1 - \ell} \\
        & = \GraphicEquationH{0.45}{0.2}{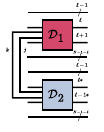} \, .
    \end{aligned}
    \label{eq:lindblad-tensor-form}
\end{equation}
Each of the terms in the odd and even layers are of this form and, therefore, commute. This allows us to derive an expression for the odd $\e^{\tau \hat{\L_o}}$ and even $\e^{\tau \hat{\L_e}}$ exponentials. Namely, using \cref{eq:dissipative-sum-two-sites,eq:lindblad-tensor-form} we can depict the tensor diagrams for $N = 4$ as
\begin{equation}
    \e^{\tau \hat{\L_o}} = \GraphicEquationH{0.45}{0.225}{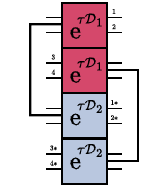} 
\end{equation}
and
\begin{equation}
    \e^{\tau \hat{\L_e}} = \GraphicEquationH{0.45}{0.225}{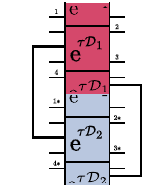} \, ,
    \label{eq:exp-odd-even-diagram}
\end{equation}
where the single thicker leg going from $e^{\tau \D_1}$ to  $e^{\tau \D_2}$ is the channel index $q$ shown in \cref{eq:kraus-two-site-tensor}. We have used \cref{eq:lk_periodic} for the PBC in $ \e^{\tau \hat{\L_e}}$, i.e., the term acting on sites 4 and 1 is shown as having half the operator on each site. A reshuffle of the legs, which we call \textit{global-to-local} transformation, allows us to bring $\D_1^{[\ell, \ell+1]}$ and $\D_2^{[\ell, \ell+1]*}$ next to each other (see \cref{def:global-to-local}). This yields the expressions in \cref{eq:exp_lindblad_even_odd}
\begin{equation*}
    \e^{\tau \hat{\L}_{o, L}} = \bigotimes_{\ell} \e^{\tau \hat{\D}^{[2\ell-1,2\ell]}} \ \text{ and } \
    \e^{\tau \hat{\L}_{e, L}} = \bigotimes_{\ell} \e^{\tau \hat{\D}^{[2\ell,2\ell+1]}} \, .
\end{equation*}

The standard ordering of the legs like the one in $\hat{\L}^{[\ell,\ell+1]}$ is based on the \textit{global} vectorization of the $N$-sites density matrix

\begin{equation}
    \begin{aligned}
        \GraphicEquationH{0.5}{0.1}{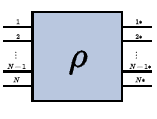} \xrightarrow{\text{Global}} \Lket{\rho} &= \GraphicEquationH{0.5}{0.1}{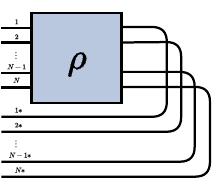} \\ 
        &= \GraphicEquationH{0.5}{0.095}{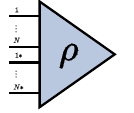} \, .
        \label{eq:rho_n_sites_vec_global}
    \end{aligned}
\end{equation}
In other words, we have sites $[1, \ldots ,N]$ followed by their conjugated counterpart $[1*, \ldots, N*]$, as shown in \cref{eq:lindblad-full-vec-diagram}. On the other hand, \cref{eq:exp_lindblad_even_odd} is based on a \textit{local} vectorization where every two sites $[\ell, \ell+1]$ are followed by their conjugate versions $[\ell*, \ell+1*]$. The local vectorization of the density matrix looks like
\begin{equation}
    \GraphicEquationH{0.5}{0.1}{rho_N_sites.pdf} \xrightarrow{\text{Local}} \Lket{\rho} = \GraphicEquationH{0.5}{0.1}{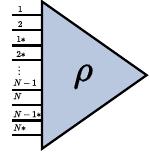} \, .
    \label{eq:rho_n_sites_vec_local}
\end{equation}

We can define a transformation that converts between these two vectorizations.

\begin{definition}[Global-to-local]\label{def:global-to-local}
    Given a multi-partite superoperator $\Lambda_G \in \fancyL (\H_1 \otimes \H_2 \ldots \otimes \H_N \otimes \H_{1*} \otimes \H_{2*} \otimes \ldots \otimes \H_{N*})$, the global-to-local transformation $G_L$ is defined as the map
    \begin{equation}
        \begin{aligned}
            G_L: \, &\fancyL (\H_1 \otimes \H_2 \ldots \otimes \H_N \otimes \H_{1*} \otimes \H_{2*} \otimes \ldots \otimes \H_{N*}) \to \\
            & \fancyL (\H_1 \otimes \H_2 \ldots \otimes \H_{N-1} \otimes \H_N \otimes \H_{N-1*} \otimes \H_{N*}) \\
            : \, & \Lambda_G \to \Lambda_L = (\Lambda_G )^{G_L} \, .
        \end{aligned}
        \label{eq:global-to-local-map}
    \end{equation}
    Applying this transformation to a superoperator $\Lambda$ is depicted as
    \begin{equation}
        \left( \GraphicEquationH{0.5}{0.1}{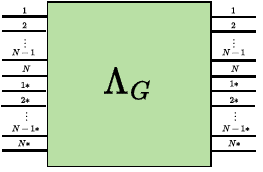} \right)^{G_L} = \GraphicEquationH{0.5}{0.1}{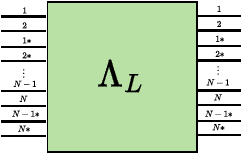} \, .
    \end{equation}
    The inverse map follows trivially, and we call it the \textit{local-to-global} transformation
    \begin{equation}
        L_G : \Lambda_L \to \Lambda_G = (\Lambda_L)^{L_G} \, .
        \label{eq:local-to-global-map}
    \end{equation}
\end{definition}

\begin{remark}
    Notice that the localized version of the superoperator arising from the global-to-local operation is defined in this way because the operators $L_k$ act on two sites. A similar construction would be obtained if instead each operator acted on $m$ sites. In this case, we would have  sites $[\ell, \ldots, \ell + m - 1]$ followed by their conjugated version $[\ell*, \ldots, \ell + m - 1*]$.
\end{remark}

The local vectorization is the natural ordering that arises from the tensor product in \cref{eq:exp_lindblad_even_odd} where each local superoperator looks like
\begin{equation}
    \e^{\tau \hat{\D}^{[\ell,\ell+1]}} = \GraphicEquationH{0.5}{0.1}{local_superoperator.pdf} .
    \label{eq:superoperator-local-diagram}
\end{equation}

Therefore, if we tensor these operators across all sites, we obtain the alternating pattern of the legs we see in \cref{eq:rho_n_sites_vec_local}. For $N = 4$, the localized odd layer diagram is
\begin{equation}
    \e^{\tau \hat{\L}_{o, L}} =   \GraphicEquationH{0.5}{0.15}{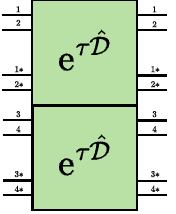} \, .
    \label{eq:exp-odd-local-diagram}
\end{equation}
The even layer follows a similar pattern by taking care of the PBC.

\section{\label{app:cholesky-decomp}Numerical Cholesky factorization}
As explained in \secref{sec:imp-ansatz}, we factorize the Choi matrix corresponding to the Lindbladian quantum channels by employing the Cholesky decomposition\cite{cholesky} of a positive-semidefinite matrix $\Choi$ into its factors $\mathcal{F}$ and $\mathcal{F}^\dagger$
\begin{equation}
    \Choi = \mathcal{F} \mathcal{F}^\dagger \, .
    \label{eq:cholesky}
\end{equation}
In practice, we achieve the Cholesky factorization through the truncated SVD decomposition
\begin{equation}
    \Choi_\Rank = V D V^\dagger = V \sqrt{D} \sqrt{D} V^\dagger = V' V'^\dagger \, ,
    \label{eq:choi-svd-decomposition}
\end{equation}
where $V$ and $V'$ are $d^4 \times \Rank$ isometries, $D$ is a $\Rank \times \Rank$ diagonal matrix, and $\sqrt{D}$ is its element-wise square root. The SVD scheme is preferred in this work over the direct Cholesky decomposition implemented in \code{NumPy}, as the latter suffers from numerical instabilities when the Choi matrix has negative eigenvalues that are zero up to numerical precision.

\section{\label{app:ranks-schemes}Comparison of ranks for different approximation schemes}

In this section, we present additional data demonstrating the increase in the Choi rank $R_N$ of the superoperator acting on all $N=4$ sites as the number of time steps $n_\tau$ increases. We provide a comparison of the ranks obtained using the Trotter, Riemannian, and SP schemes, which were introduced in \secref{sec:sim-alg-comparison}. The comparison for the Kitaev model with $\tau=1$ and $\Rank = 2$ is shown in \cref{tab:app-ranks-kitaev}.
\begin{table*}[htb]
    \centering
    \renewcommand{\arraystretch}{1.5} 
    \begin{tabular}{c | c| c| c| c| c}
    \hline\hline
    $n_\tau$ & Trotter & Riemannian & SP 1st order & SP 2nd order & SP 3rd order \\
    \hline
    1 & 36 & 46 & 5 & 19 & 29 \\
    2 & 45 & 68 & 19 & 45 & 45 \\
    3 & 45 & 87 & 45 & 45 & 45 \\
    4 & 45 & 130 & 45 & 45 & 45 \\
    5 & 45 & 160 & 45 & 45 & 45 \\
    \hline
    \end{tabular}
    \caption{Comparison of full Choi ranks for different approximation schemes using the Kitaev model for increasing number of time steps.}
    \label{tab:app-ranks-kitaev}
\end{table*}

Analogously, in \cref{tab:app-ranks-pspl}, we show the comparison of the ranks when using the PSPL model with $\tau=1$ and $\Rank = 10$.

\begin{table*}[htb]
    \centering
    \renewcommand{\arraystretch}{1.5}
    \begin{tabular}{c | c| c| c| c| c| c}
    \hline\hline
    $n_\tau$ & Trotter & Riemannian & SP 1st order & SP 2nd order & SP 3rd order & SP 4th order \\
    \hline
    1 & 256 & 256 & 7 & 38 & 96 & 190 \\
    2 & 256 & 256 & 38 & 241 & 256 & 256 \\
    3 & 256 & 256 & 145 & 256 & 256 & 256 \\
    4 & 256 & 256 & 241 & 256 & 256 & 256 \\
    5 & 256 & - & 256 & 256 & 256 & 256\\
    \hline
    \end{tabular}
    \caption{Comparison of full Choi ranks for different approximation schemes using the PSPL model for increasing number of time steps.}
    \label{tab:app-ranks-pspl}
\end{table*}

\section{Parametrizing the tangent space}\label{app:imp-parametrization-sm}

The choice of the Riemannian metric, inducing an inner product on the tangent space, determines the form of the Riemannian elements. Thus, it might impact the performance of the optimization algorithm. Therefore, in the next section, we discuss the use of different tangent space metrics. To this end, let us start this section by following the procedure outlined in \cite{unitary_constraints} to derive parametrized definitions for the tangent and normal spaces of the Stiefel manifold that will come in handy both for the metric discussions and the numerical implementations. We start by looking at the projection operator $\pi: \R^{n \times p} \to \St$. This operation acts on an arbitrary rank-$p$ matrix $Y \in \R^{n \times p}$ and projects it onto the closest point on $\St$
\begin{equation}
    \pi (Y) = \argmin_{X \in \St} \norm*{Y - X}^2 \, ,
    \label{eq:imp-projection-general}
\end{equation}
where $\norm*{.}$ is the Frobenius norm induced by the Euclidean inner product. Let us now look at the perturbed point $\pi(X + \epsilon Y)$, for $X \in \St$, $Y \in \R^{n \times p}$, and $\epsilon \in \R$ a scalar. Since the Stiefel manifold does not fill the entire $\R^{n \times p}$ space, there are some directions $Y$ that satisfy
\begin{equation}
    \pi(X + \epsilon Y) = X + \O(\epsilon^2) \, .
    \label{eq:imp-normal-definition}
\end{equation}
In other words, moving in the direction of $Y$ does not take $\pi(X + \epsilon Y)$ away from $X$. We recognize the collection of such directions $Y$ as the \textit{normal space}. Using this practical definition, the tangent space can be defined as the orthogonal complement of the normal space.
Now, to find the concrete parametrization of these spaces, let us first define the column orthogonal complement of the isometry $X \in \St$ as the matrix $\Xperp \in \R^{n \times (n-p)}$, satisfying 
\begin{align}
    [X \Xperp]^T[X \Xperp] &= \I_n \label{eq:imp-orthogonal} \\
    X X^T + \Xperp \Xperp^T &= \I_n \, , \label{eq:imp-orthogonal-4}
\end{align}
where $\I_n$ is the $n \times n$ identity matrix and the side-by-side stacking of matrices $[A B]$ is a block-matrix yielding a row vector with two matrix elements $\left( A, B \right)$. Analogously, the column vector $[A B]^T$ arises from the vertical stacking of the matrices. We can use these relations to define the decomposition of an arbitrary matrix $Y \in \R^{n \times p}$
\begin{equation}
    Y = XA + \Xperp B + X C
    \label{eq:imp-decomposition-x-x-perp}
\end{equation}
in terms of a skew-symmetric matrix $A \in \R^{p \times p}$, a symmetric matrix $C \in \R^{p \times p}$, and an arbitrary matrix $B \in \R^{(n-p) \times p}$. This important decomposition is used in Lemma 8 of \cite{unitary_constraints} to prove the relation
\begin{equation}
    \pi(X + \epsilon Y) = X + \epsilon(X A + \Xperp B) + \O(\epsilon^2) \, .
    \label{eq:imp-tangent-decomposition}
\end{equation}
It then follows that using $Y = XC$ we get $\pi(X + \epsilon XC) = X + \O(\epsilon^2)$, and for a small enough $\abs*{\epsilon}>0$, this yields
\begin{equation}
    \pi(X + \epsilon XC) = X \, .
    \label{eq:imp-normal-derivation}
\end{equation}
We recognize this expression as the definition of the normal space.
\begin{definition}[Normal space]\label{def:imp-normal-space-sm}
    Given an element $X \in \St$ of the Stiefel manifold, the normal space $N_X \St$ at $X$ is
    \begin{equation}
        N_x \St = \{ N \in \R^{n \times p}: N = XC, C \in \R^{p \times p}, C=C^T \} \, .
        \label{eq:imp-normal-space-param-sm}
    \end{equation}
\end{definition}
Inspired by \cref{eq:imp-tangent-decomposition}, we can infer the form of the tangent space vectors to be $XA + \Xperp B$, which yields the parametrized definition of the tangent space.

\begin{definition}[Tangent space - revisited]\label{def:imp-tangent-space-sm}
    Given an element $X \in \St$ of the Stiefel manifold, the tangent space $T_X \St$ at $X$ is
    \begin{equation}\label{eq:imp-tangent-space-param-sm}
        \begin{aligned}
            T_X \St &= \{ Z \in \R^{n \times p}: Z = XA + \Xperp B, A \in \R^{p \times p}, \\
            & A + A^T = 0, B \in \R^{(n-p)\times p} \} \, .
        \end{aligned}
    \end{equation}
    If we compare \cref{eq:imp-tangent-space-param-sm,eq:tangent-stiefel}, we realize the product $X^T Z$ in the former is exactly the skew-symmetric matrix $A$ of the latter, for $Z \in T_X \St$.
\end{definition}
It is straightforward to verify that $XA + \Xperp B$ is orthogonal to $XC$ under the Euclidean inner product
\begin{equation}
    \Inner{XA + \Xperp B}{XC}_e = \trace \{ (XA + \Xperp B)^T XC \} = 0 \, ,
    \label{eq:imp-orthogonal-tangent-normal}
\end{equation}
where we used $\tr(A^T C) = 0$. 
From the revisited definitions of the tangent and normal space, we can verify that at $X \in \St$
\begin{equation}
    \pi_{T}(Y) = (I - X X^T) Y + X \Skew(X^T Y)
    \label{eq:imp-tangent-projector-stiefel}
\end{equation}
is the projection of an arbitrary matrix $Y \in \R^{n \times p}$ onto the tangent space with the $\Skew$ operator 
\begin{equation}
    \Skew(M) = \frac{1}{2}(M -M^T) \, .
    \label{eq:imp-skew}
\end{equation}
The first term comes from using \cref{eq:imp-orthogonal-4,eq:imp-orthogonal,eq:imp-decomposition-x-x-perp} as
\begin{equation}
    (I - X X^T)Y = \Xperp \Xperp^T Y = \Xperp B \, .
    \label{eq:imp-b-factor}
\end{equation}
The second term uses \cref{eq:imp-orthogonal,eq:imp-decomposition-x-x-perp} to obtain $X^T Y = A + C$, followed by the $\Skew$ operator defined in \cref{eq:imp-skew} to ``pick'' the skew-symmetric factor
\begin{equation}
    X \Skew(X^T Y) = XA \, .
    \label{eq:imp-a-factor}
\end{equation}
Analogously, the projection onto the normal space is
\begin{equation}
    \pi_{N}(Y) = X \, \Symm(X^T Y) = \frac{X}{2}(Y^T X + X^T Y)\, .
    \label{eq:imp-normal-space}
\end{equation}
using the symmetric operator
\begin{equation}
    \Symm{A} = \frac{1}{2}(A + A^T)
    \label{eq:imp-symm}
\end{equation}
that picks the symmetric factor $\Symm(X^T Y) = C$. Finally, we multiply by $X$, yielding the parametrization in \cref{eq:imp-normal-space-param-sm}. We can simplify \cref{eq:imp-tangent-projector-stiefel} to obtain a relation between $\pi_{N}$ and $\pi_{T}$
\begin{equation}
        \begin{aligned}
            \pi_{T}(Y) & = Y - \frac{X}{2}(Y^T X + X^T Y )\\
            & = Y - \pi_{N}(Y) \, .
        \end{aligned}
    \label{eq:imp-tangent-normal-relation}
\end{equation} 

\section{Riemannian connection and metric\label{app:connection-metric}}

As we saw \cref{eq:hessian-riemannian}, the Riemannian connection is a fundamental element in our second-order optimization scheme. Here, we describe the relation between the Riemannian connection $\nabla$ and the Riemannian metric $G_X$. Let $\overline{\nabla}$ and $\nabla$ be the Riemannian connections of the Euclidean space $\E$ and the Stiefel manifold $\St$, respectively. By proposition 5.3.2 of \cite{princeton_manifolds} we obtain the relation
\begin{equation}
    \nabla_{W} \hat{Z} = \pi_{T} \left( \overline{\nabla}_{W}\hat{Z} \right) \, ,
    \label{eq:riemannian-connection}
\end{equation}
for all $W \in \TxSt{X}$ and $\hat{Z} \in \SVF(X)$. 
The ambient Riemannian connection $\overline{\nabla}$ is given in Theorem 3.1 of \cite{nguyenoperatorvalued_metric_based} as
\begin{equation}
    \overline{\nabla}_{W} \hat{Z} := D \hat{Z}(X)[W] + G^{-1}_X \K(W, \hat{Z})
    \label{eq:ambient-riemannian-connection}
\end{equation}
in terms of the differential map (a classical directional derivative)
\begin{equation}
    D \hat{Z}(X)[W] = \lim_{t \to 0} \frac{\hat{Z}(X + t W) - \hat{Z}(X)}{t} \, ,
    \label{eq:tangent-map}
\end{equation}
and the Christoffel metric term $\K \in \E$
\begin{equation}
    \K(W, \hat{Z}) := \frac{1}{2}\left[(D G_X [W])\hat{Z} + (D G_X [\hat{Z}])W - \X(W, \hat{Z})\right] \, .
    \label{eq:christoffel-term}
\end{equation}
The cross term $\X(W, \hat{Z}) \in \E$ in this expression is a bilinear form such that for any vector field $\hat{Z}_0$
\begin{equation}
    \Inner{\X(W, \hat{Z})}{\hat{Z}_0}_e = \Inner{W}{(D G_X[\hat{Z}_0])\hat{Z}}_e  \, ,
    \label{eq:cross-term}
\end{equation}

\section{Choosing a Riemannian metric}\label{app:imp-metric}

Let us employ the parametrization and dimension counting tools we have just described in \secref{app:imp-parametrization-sm} for the discussion of the Riemannian metric choice. The Euclidean metric $\Inner{.}{.}_e$ we have used thus far seems like the natural choice for a submanifold of the Euclidean space. However, it turns out this inner product does not weigh the degrees of freedom of the Stiefel tangent space equally. To see this, let us look at the Euclidean inner product of $Z \in T_X \St$
\begin{align}
    \Inner{Z}{Z}_e &= \tr A^T A + \tr B^T B = 2 \sum_{i < j} a_{i j}^2 + \sum_{i, j} b_{i j}^2 \, ,
    \label{eq:imp-inner-ununiformly}
\end{align}
where we have summed only over the upper diagonal for $A$. We can see from this equation that the Euclidean metric counts the degrees of freedom of $A$ twice as much as the ones for $B$. As shown in \cite{edelman1998geometry_stiefel}, an alternative metric is the \textit{canonical metric} defined for $Z, W \in T_X \St$ as
\begin{equation}
    \Inner{Z}{W}_c = \tr { Z^T (\I - \frac{1}{2}X X^T) W}.
    \label{eq:imp-canonical-metric}
\end{equation}
Using this metric, the independent coordinates of $A$ and $B$ are weighted equally at every $X \in \St$
\begin{align}
    \Inner{Z}{Z}_c &= \frac{1}{2}\tr A^T A + \tr B^T B = \sum_{i < j} a_{i j}^2 + \sum_{i, j} b_{i j}^2 \, .
    \label{eq:imp-canonical-uniformly}
\end{align}
So far, we have studied the Euclidean and canonical metrics. However, it turns out these are not the only inner products that induce a Riemannian metric on the Stiefel manifold. In \cite{nguyenoperatorvalued_metric_based}, a general formulation for the action of the ambient metric $G_X$ at $X \in \St$ is given by
\begin{equation}
    G_X Z = \alpha_0 (\I - X X^T)Z + \alpha_1 X X^T Z\, .
    \label{eq:imp-g-metric}
\end{equation}
For $\alpha_0, \alpha_1 > 0$, this metric induces an inner product in $T_X \St$, and hence it induces a Riemannian metric in $\St$. The inverse matrix is
\begin{equation}
    G_X^{-1} Z = \frac{1}{\alpha_0}(\I - X X^T)Z + \frac{1}{\alpha_1} X X^T Z.
    \label{eq:imp-g-inverse}
\end{equation}
Then, we conclude the metrics described thus far are special cases of the general formulation in \cref{eq:imp-g-metric}, with $\alpha_0 = \alpha_1 = 1$ and $\alpha_0 = 1$, $\alpha_1 = 1/2$ for the Euclidean and the canonical metric, respectively.
As anticipated, the specific form of the gradient depends on the metric of choice. Using \cref{eq:imp-tangent-projector-stiefel,eq:imp-g-inverse} we can rewrite the Riemannian gradient expression in \cref{eq:gradient-metric} as
\begin{equation}
    \begin{aligned}
        \gradR f(x) &= \pi_T \left( G^{-1} \gradR \overline{f}(x)\right) \\
        &= \frac{\gradR \overline{f}(x)}{\alpha_0} + \frac{\alpha_1^{-1} - 2\alpha_0^{-1}}{2} X X^T \gradR \overline{f}(x) \\
        &- \frac{X \gradR \overline{f}(x)^T X}{2 \alpha_1} \, .
    \end{aligned}
    \label{eq:imp-gradient-metric}
\end{equation}

As shown in \cref{eq:riemannian-connection}, the metric choice also affects the Riemannian connection and, with it, the Riemannian Hessian calculation. We can use \cref{eq:imp-g-metric} to obtain the cross-term in \cref{eq:christoffel-term} for a point $X \in \St$ and tangent vectors $Z, W \in T_X \St$ \cite{nguyenoperatorvalued_metric_based}

\begin{equation}
    \X(Z, W) = (\alpha_1 - \alpha_0)(Z W^T + W Z^T) X \, .
    \label{eq:imp-christoffel-term}
\end{equation}
Then, using \cref{eq:riemannian-connection,eq:ambient-riemannian-connection,eq:christoffel-term}, the general connection for the Stiefel manifold is given by \cite{nguyenoperatorvalued_metric_based}
\begin{equation}
    \begin{aligned}
        \nabla_{Z} W &= DW[Z] + \frac{1}{2} X (W^T Z + Z^T W) \\
        &+ \frac{\alpha_0 - \alpha_1}{\alpha_0}(\I - XX^T)(W Z^T + Z W^T) X \, .
    \end{aligned}
    \label{eq:imp-riemannian-connection}
\end{equation}

\section{Product of isometry manifolds}\label{app:imp-product}

In \cref{eq:cost-function-xvec}, we established the main cost function we seek to minimize in this work, namely
\begin{equation*}
   f(\Xvec) = \norm*{ \e^{\tau \hat{\L}} - \Super(\Xvec)} \, .
\end{equation*}
Since this function depends on $m$ isometries, its domain is not a single Stiefel manifold, but rather a product manifold: $\Stnp{n_1}{p_1} \times \Stnp{n_2}{p_2} \ldots \times \Stnp{n_m}{p_m}$. We will assume all the individual manifolds are of the same dimensions and label the Cartesian product as $\St^{\times m}$. We can then define the cost function $f$ as the map acting on this product manifold
\begin{equation}
    f: \St^{\times m} \to \R : f(\Xvec) \to  \norm*{ \e^{\tau \hat{\L}} - \Super(\Xvec)} \, ,
    \label{eq:imp-cost-function-product}
\end{equation}
with $\Xvec \in \St^{\times m}$. The tangent space of a product manifold is, in general, the Cartesian product of the individual tangent spaces, but for our finite dimensional tensors, it reduces to the direct sum of them \cite{Hauru_2021}. We will see what this means in practice for our Riemannian gradient, Hessian, and retraction.

\section{An appropriate retraction}\label{app:imp-retraction}

Back in \cref{def:retraction}, we introduced another of the cornerstone concepts we employ in our optimization algorithm: the retraction $R_X$. This is used within the trust-region algorithm to define the first-order model in \cref{eq:model-first-order} and to generate the next candidate at the $\beta$-th step in the optimization algorithm $X^{\beta+1} = R_{X^\beta}(Z)$, with $Z$ the solution of the inner iteration in \cref{eq:trust-subproblem}. As discussed in \secref{app:imp-product}, the tangent space $T_\Xvec \St^{\times m}$ of the product manifold $\St^{\times m}$ results from the direct sum of the individual tangent spaces. Therefore, to implement a retraction from $\Zvec \in T_\Xvec \St^{\times m}$ onto $\Xvec \in \St^{\times m}$, all we need is to apply the retraction to each pair $(X_{\alpha} \in \Xvec, Z_\alpha \in \Zvec)$ individually. The first-order retraction for the Stiefel manifold used in this work is the \textit{canonical generalized polar decomposition} for any rectangular matrix $M \in \R^{n \times p}$
\begin{equation}
    M = X C \, ,
    \label{eq:imp-canonical-polar}
\end{equation}
in terms of an isometry $X \in \R^{n \times p}$  and a positive semidefinite matrix  $C \in \R^{p \times p}$ \cite{canonical-polar}. Given the SVD decomposition of $M$,

\begin{equation*}
    M = U D V^T \, ,
\end{equation*}
the isometry factor $X$ is
\begin{equation}
    X = U \, \I_{n,p} \,  V^T \, ,
    \label{eq:imp-svd-polar}
\end{equation}
with $\I_{n, p}$ the first $p$ columns of the $n \times n$ identity matrix.

\section{Building the Riemannian gradient}\label{app:imp-gradient}

A central element for the trust-region method used in \cref{alg:riemannian-approach-main} is the Riemannian gradient of the function $f$ from \cref{eq:imp-cost-function-product}. As mentioned in \secref{app:imp-product}, the tangent space of the product manifold is the direct sum of the individual tangent spaces. In practice, this means we can treat the tangent vectors $Z$, living on the tangent space $\TxStnp{X}{n}{p}$ at $X \in \St$, as independent of the rest and concatenate them to obtain $\Zvec \in \TxSt{\Xvec}^{\times m}$. Based on the arguments developed in \secref{app:imp-metric}, we will employ the equitative canonical metric in this work. Consequently, the Riemannian gradient for every $X \in \Xvec$ takes the form
\begin{equation}
    \gradR f(X) =\gradR \overline{f}(x) +  X \gradR \overline{f}(x)^T X \, .
    \label{eq:imp-gradient-canonical}
\end{equation}
The ambient gradient $\gradR \overline{f}(X)$ coincides with the Euclidean gradient in the above equation. As aforementioned, to obtain $\gradR f(\Xvec) \in \TxSt{\Xvec}^{\times m}$, we will concatenate the $m$ individual Riemannian gradients; in other words, we will \textit{stack} them.

After using the ambient gradients together with \cref{eq:imp-gradient-canonical} to obtain the Riemannian gradients, a further intermediate step is the parametrization of the tangent vectors. As we saw in \cref{eq:imp-dof-ansatz}, the tangent spaces $\TxStnp{X}{n}{p}$ are parametrized by
\begin{equation*}
    \DOF = \frac{m p}{2}\left( 2n - p - 1 \right) = mnp - \frac{m p}{2}\left(p + 1\right)\
\end{equation*}
degrees of freedom. Therefore, we can store and manipulate $mnp - \frac{m p}{2}(p + 1)$ parameters, instead of $mnp$. To achieve this, we use \cref{eq:imp-a-factor,eq:imp-b-factor} to obtain $A$ and $B$ for each $\gradR f(X)$, respectively, retrieving the $p(p-1)/2$ upper triangular elements of $A$ and stacking them on top of all $p(n-p)$ elements of $B$. For convenience, we reshape all the stacked parametrized gradients into a vector of length $\DOF$.

\section{Building the Riemannian Hessian}\label{app:imp-hessian}
The last missing piece in our optimization algorithm is the calculation of the Riemannian Hessian. As we saw in \cref{eq:model-first-order}, this linear mapping from $T_X \St$ to $T_X \St$ allows us to define an approximation of the cost function $f$, which makes it a fundamental element of the trust-region algorithm. 

We have discussed the general form of the Riemannian connection of the Stiefel manifold in \cref{eq:imp-riemannian-connection}. Using the canonical metric $G_X^{[1, 1/2]}$, this becomes \cite{nguyenoperatorvalued_metric_based}

\begin{equation}
    \begin{aligned}
        \nabla_{Z} W &= DW[Z] + \frac{1}{2} X (W^T Z + Z^T W) \\
        & + \frac{1}{2}(\I - XX^T)(W Z^T + Z W^T) X \, .    
    \end{aligned}
\end{equation}
Together with the Riemannian Hessian-vector product for $Z \in T_X \St$ defined as
\begin{equation}
    \Hess f(X)[Z] = \nabla_{Z} \gradR f (X) \, ,
\end{equation}
we obtain the relation
\begin{equation}\label{eq:imp-hessian-connection}
    \begin{aligned}
        & \Hess f(X)[Z] = D \gradR f (X) [Z] \\
        & + \frac{1}{2} X ( \gradR f (X) ^T Z + Z^T  \gradR f (X) )\\
        & + \frac{1}{2}(\I - XX^T)( \gradR f (X)  Z^T + Z  \gradR f (X) ^T) X \, .
    \end{aligned}
\end{equation}
In order to compute the full Riemannian Hessian $\Hess f(X)$, we take inspiration from the Euclidean counterpart, where each column of the Euclidean Hessian $H_f(x)_k$ is revealed by the Hessian-vector product with the basis vectors $e_k \in \R^{n}$
\begin{equation*}
    H_f(x)_k  = H_f (x) e_k \, .
\end{equation*}
Here, we will follow a similar approach, and generalize the concept of a basis vector on $\R^n$ by means of the coordinate vector fields or elementary tangent directions on the tangent space of a matrix manifold \cite{princeton_manifolds}. Using the canonical metric, the elementary tangent directions are \cite{unitary_constraints}:
\begin{alignat}{2}
    E^A_{ij} &= X (E_{ij} - E_{ji}), && \quad E_{ij}, E_{ji} \in \R^{p \times p}\label{eq:imp-elementary-anti}  \, , \\
    E^B_{ij} &=  \Xperp E_{ij}, && \quad E_{ij} \in \R^{(n-p) \times p}  \label{eq:imp-elementary-arbitrary} \, ,
\end{alignat}
where $E_{ij}$ is a unit matrix defined as one at the element $(i,j)$ and zero everywhere else, for appropriate $(i,j)$ index values. We notice \cref{eq:imp-elementary-anti} corresponds to the antisymmetric component $XA$ of the tangent vectors, so $(i,j)$ should traverse the $p(p-1)/2$ upper diagonal elements. On the other hand, \cref{eq:imp-elementary-arbitrary} corresponds to the arbitrary component $\Xperp B$, so the indices $(i,j)$ traverse all $p(n-p)$ values. We define the flattened index $\mathrm{idx} \in \dims{p\left( 2n - p - 1 \right)/2}$, and the corresponding elementary direction $E^{[\mathrm{idx}]}$, such that
\begin{equation}
    E^{[\mathrm{idx}]} =
    \begin{cases}
        E^A_{ij} & \text{if } \mathrm{idx} \leq \frac{p(p-1)}{2} \\
        E^B_{ij} & \text{else}
    \end{cases} \, .
    \label{eq:elementary-direction-index}
\end{equation}
Thus, for a single isometry $X$, we obtain the relation
\begin{equation}
    \Hess f(X)_{\mathrm{idx}} = \Hess f(X)[E^{[\mathrm{idx}]}] \, .
    \label{eq:imp-hessian-idx-elementary}
\end{equation}
The classical directional derivative $D \gradR f(X) [Z]$ of the vector-valued function $\gradR f(X)$ is implemented through the Jacobian-vector product. Once again, we take advantage of the automatic differentiation capabilities of \code{JAX}, and in particular we employ the forward-mode differentiation function \code{jvp} \cite{jax2018github}.

In order to accommodate this procedure to the product manifold structure, we compute the  $(\mathrm{idx}, \alpha)$ element of the Riemannian Hessian $\Hess f(\Xvec)$ using the tangent directions constructed as the $m$-length list
\begin{equation}
    \text{tangents} = \left[ O_{n \times p}, \ldots , E^{[idx]}, \ldots, O_{n \times p}  \right] \, ,
    \label{eq:imp-tangents-list}
\end{equation}
where the only non-zero matrix is the $\alpha$-th element $E^{[\mathrm{idx}]}$. Then, each of the resulting elements from \code{jax.jvp} is projected using the Riemannian connection in \cref{eq:imp-riemannian-connection} to obtain the corresponding tangent vector in $T_\Xvec \St$. We store the corresponding parametrization on $\Hess f(\Xvec)_{\mathrm{idx}, \alpha}$. To obtain the full Riemannian Hessian, we repeat the process for all $\alpha \in \dims{m}$ and $\mathrm{idx} \in \dims{p\left( 2n - p - 1 \right)/2}$, and reshape the Riemannian Hessian into a $\DOF \times  \DOF$ dimensional matrix.

\section{\label{app:sim-parametrization}The importance of a good parametrization}

In \secref{app:imp-parametrization-sm}, we discussed how to parametrize the tangent spaces of the Stiefel manifold. Later, in \secref{app:imp-metric}, we used these ideas to derive expressions that showcase the influence of the Riemannian metric in calculating the gradient and Hessian. We now explore how the choice of the metric affects the performance of the Riemannian optimization in practice.

\begin{figure}[t]
    \begin{subfigure}{0.45\textwidth}
        \includegraphics[width=1\textwidth]{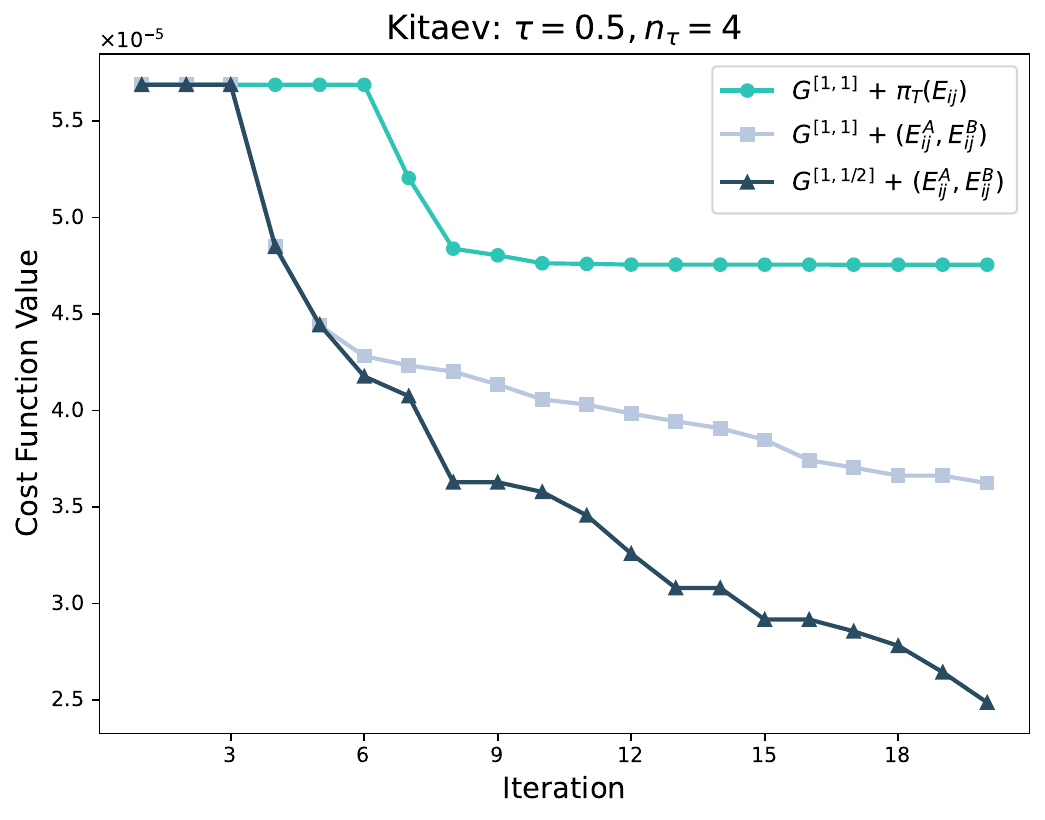}
        \caption{Optimization under Kitaev model.}
        \label{fig:kitaev-param-metric}
    \end{subfigure}
    \begin{subfigure}{0.45\textwidth}
        \includegraphics[width=1\textwidth]{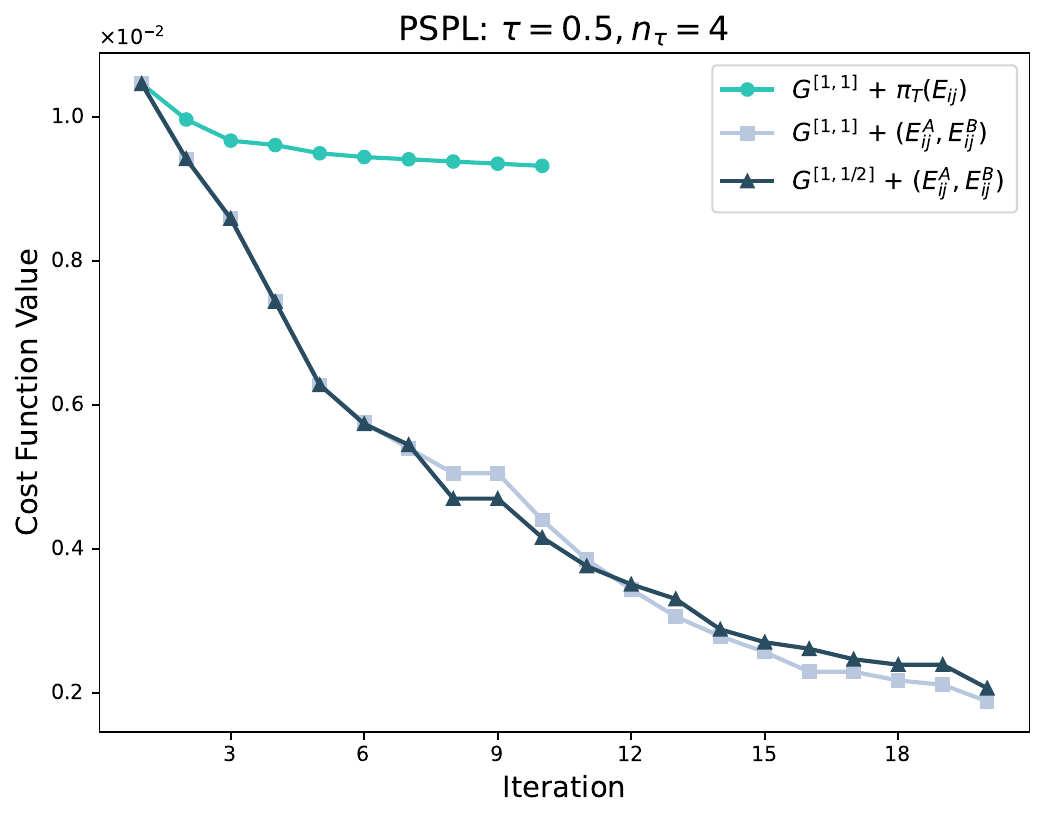}
        \caption{Optimization under PSPL model.}
        \label{fig:pspl-param-metric}
    \end{subfigure}
    \caption{Comparison of the cost function for a final evolution time of $\tau = 0.5$ and $n_\tau=4$ time steps, after 20 optimization iterations using different combinations of metric $G$ and elementary tangent directions $E_{i,j}$ (see main text). The \textit{dark blue} curve shows the choice used in this work. (a) Optimization for the Kitaev model. (b) Optimization for the PSPL model. The curve with $\pi_T(E_{ij})$ is only shown for 10 time steps as it reaches a saddle point early on.}
\end{figure}

To this end, we will carry our optimization scheme for three different combinations of metric and elementary tangent directions:

\begin{itemize}
    \item Euclidean metric $G^{[1,1]}$ and a suboptimal elementary tangent direction $\pi_T(E_{ij})$, i.e., the unit matrix $E_{ij}$ projected onto the tangent space.
    \item Euclidean metric $G^{[1,1]}$ and the elementary tangent directions $(E^A_{ij}, E^B_{ij})$ introduced in \cref{eq:imp-elementary-anti,eq:imp-elementary-arbitrary}.
    \item Canonical metric $G^{[1,1/2]}$ and elementary tangent directions $(E^A_{ij}, E^B_{ij})$.
\end{itemize}
\cref{fig:kitaev-param-metric} shows the value of the cost function $f(\Xvec^{\beta})$ for 20 iterations under the Kitaev model with $n_\tau=4$ time steps and evolution time of $\tau = 0.5$. This comparison validates the choice of the canonical metric as the Riemannian metric of the Stiefel manifold since it clearly outperforms the optimization using the Euclidean counterpart. This difference, as argued in \secref{app:imp-metric}, can be attributed to the correct weight assigned by the canonical metric to each degree of freedom parametrizing the tangent space vectors. On the contrary, the Euclidean metric assigns twice as much weight to the antisymmetric components, as shown in \cref{eq:imp-inner-ununiformly}. In addition, the poor performance of the elementary directions $\pi_T(E_{ij})$, shown by the turquoise curve, confirms the importance of choosing properly how to traverse the tangent space when creating the Riemannian Hessian using \cref{eq:imp-hessian-idx-elementary}. In \cref{fig:kitaev-param-metric}, we also see that using this elementary direction results in a fast convergence to a suboptimal local minimum.

However, it turns out the canonical metric is not always the obvious choice. \cref{fig:pspl-param-metric} shows the cost function optimization under the PSPL model and final evolution time $\tau=0.5$ with a time step $n_\tau=4$. Contrary to the results from \cref{fig:kitaev-param-metric}, the comparison in \cref{fig:pspl-param-metric} shows that the Euclidean and canonical metrics yield very similar results when using the same elementary tangent directions. Moreover, the Euclidean metric even outperforms the canonical metric at some intervals of the iterations.

\section{\label{app:sim-convergence-layers}Expressivity vs. iterations}

In \secref{sec:sim-rank-up}, we carried out the Riemannian optimization using $n_\tau=1$ and $n_\tau=4$ as the number of time steps for the PSPL and Kitaev models, respectively. However, an important question to ask when working with iterative optimization methods is \textit{``How many iterations do we need for the algorithm to converge to a minimum''}. In this section, we will try to address this question numerically by looking at the behavior of the cost function when increasing the number of time steps. This dependence is particularly interesting to us, as the degrees of freedom in the optimization grows linearly with the number of time steps $n_\tau$, i.e., the number of time steps increases the expressivity.

\begin{figure}[t]
    \begin{subfigure}{0.45\textwidth}
        \includegraphics[width=0.925\textwidth]{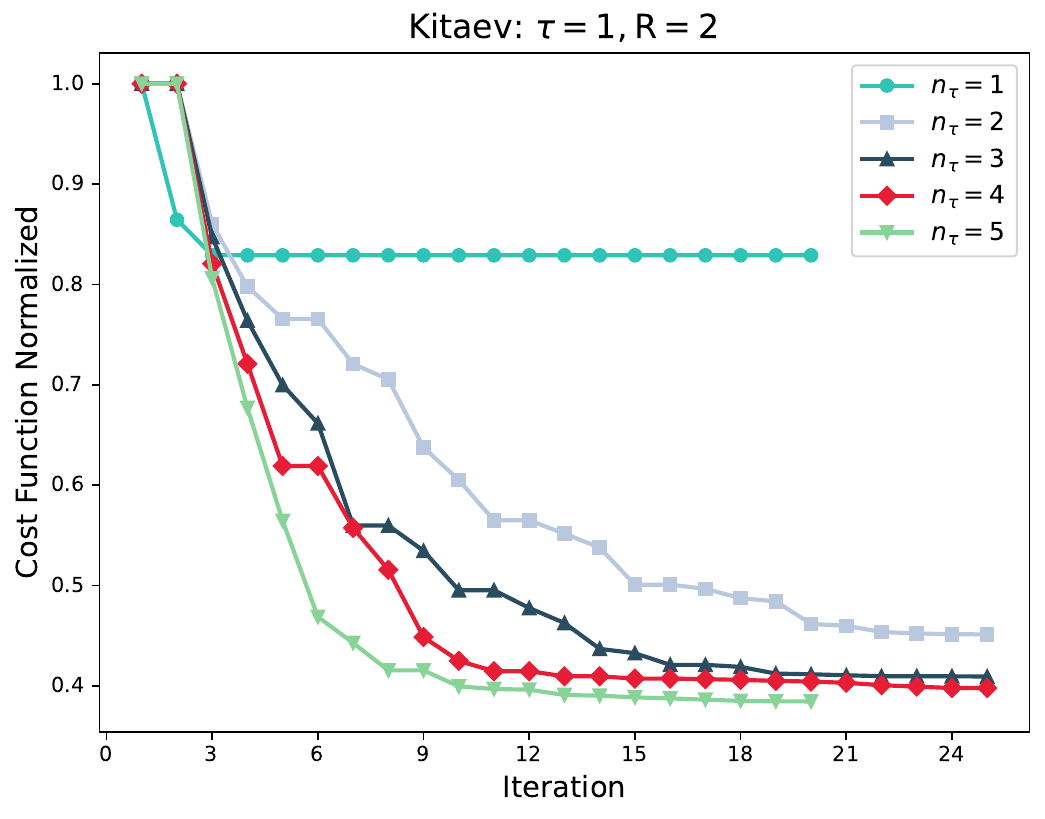}
        \caption{Normalized approximation error for the Kitaev model, with $\tau = 1$ and $\Rank = 2$.}
        \label{fig:sim-convergence-kitaev}
     \end{subfigure}
     \begin{subfigure}{0.45\textwidth}
        \includegraphics[width=0.965\textwidth]{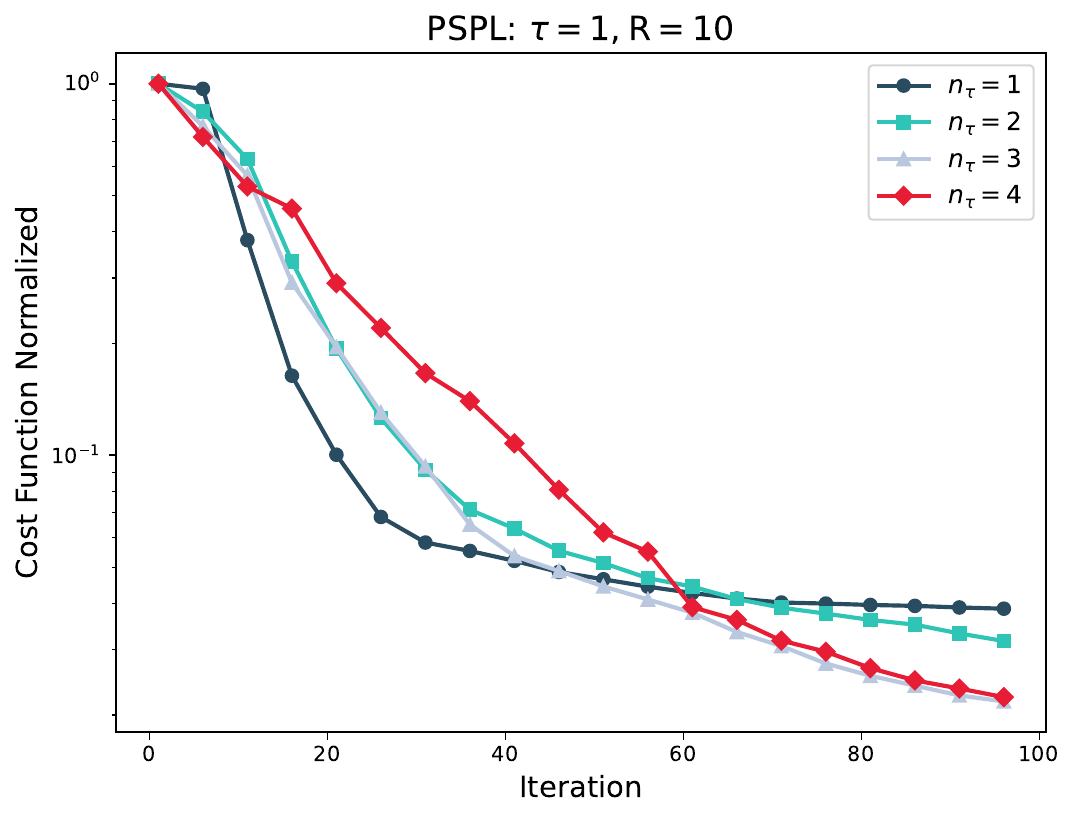}
        \caption{Normalized approximation error for the PSPL model, with $\tau = 1$ and $\Rank = 10$.}
        \label{fig:sim-convergence-pspl}
     \end{subfigure}
     \caption{Normalized approximation error as a function of the number of iterations for both models, while varying the number of time steps $n_\tau$.}
\end{figure}

 First, we look at the optimization with the Kitaev model, using $\tau = 1$ and $\Rank = 2$ (its natural Choi rank), and vary the number of time steps $n_\tau$. In order to compare the decrease in the cost function for each value of $n_\tau$, we look at the \textit{normalized} cost function trajectory
 \begin{equation}
    f_{\mathrm{Norm}}\left( \Xvec^\beta \right) =  \frac{f\left(\Xvec^\beta\right)}{f\left(\Xvec^0\right)} \,.
    \label{eq:sim-normalized-cost}
 \end{equation}
 The results shown in \cref{fig:sim-convergence-kitaev} provide insight to answer the question we posed at the beginning of this section. First, we notice that by increasing the number of time steps, thus increasing the expressivity, we can achieve a lower relative approximation error. We also observe that all the curves converge after less than 30 iterations. The curve with $n_\tau=1$ shows the least improvement, as it reaches a plateau after only 5 iterations. While the curves for $n_\tau=3,4,5$ all converge around the same point, surprisingly, the optimization using $n_\tau=2$ takes the longest to settle.

We repeat an analogous simulation using the PSPL model, $\tau = 1$, and $\Rank = 10$. Compared to the Kitaev scenario, the trajectories in \cref{fig:sim-convergence-pspl} show that only the optimization using the lowest number of time steps $n_\tau=1$ converges under 100 iterations. Even after decreasing the cost function over an order of magnitude, the curves for $n_\tau = 3$ and $n_\tau=4$ still depict a steep descent towards the final iterations. We can attribute this behavior to the big parameter space induced by the high-rank value $\Rank = 10$, compared to the Kitaev rank $\Rank = 2$, causing an increase in the complexity of the optimization. By a similar argument, we observe how the red curve corresponding to $n_\tau=4$ exhibits the worse performance up to the $60$-th iteration, as it requires more iterations to find a minimum.

\end{document}